\documentclass{ieeeaccess}
\usepackage{cite}
\usepackage{amsmath,amssymb,amsfonts}
\usepackage{algorithmic}
\usepackage[usenames,dvipsnames]{xcolor}
\usepackage{pgfplots}
\usepackage{orcidlink}
\usetikzlibrary{shapes.geometric}
\usetikzlibrary{positioning}
\usetikzlibrary{quantikz2}
\usetikzlibrary{shapes.misc}
\tikzset{cross/.style={cross out, draw=black, minimum size=2*(#1-\pgflinewidth), inner sep=0pt, outer sep=0pt},
cross/.default={9pt}}
\usepackage{romannum}
\usepackage{graphicx}
\usepackage[font={sf,scriptsize},           % new
            labelfont={bf,color=accessblue},% new
            caption=false]                  % new
           {subfig} 
\usetikzlibrary{arrows.meta,arrows}
\usepackage{circuitikz}
\usetikzlibrary{decorations.pathmorphing}
\usepackage{pgfplotstable}
\pgfplotsset{compat=1.7}
\usepackage{tikz}
 \NewSpotColorSpace{PANTONE}
  \AddSpotColor{PANTONE} {PANTONE3015C} {PANTONE\SpotSpace 3015\SpotSpace C} {1 0.3 0 0.2}
  \SetPageColorSpace{PANTONE}
\usepackage{graphicx}
\usepackage{textcomp}
\def\BibTeX{{\rm B\kern-.05em{\sc i\kern-.025em b}\kern-.08em
    T\kern-.1667em\lower.7ex\hbox{E}\kern-.125emX}}
\begin{document}
\history{Date of publication xxxx 00, 0000, date of current version xxxx 00, 0000.}
\doi{10.1109/ACCESS.2023.0322000}

\title{Fault-tolerant one-way noiseless amplification for microwave bosonic quantum information processing }
\author{\uppercase{Hany Khalifa}\authorrefmark{1} \orcidlink{0000-0002-1276-5428}, 
\uppercase{Riku J\"{a}ntti}\authorrefmark{1} \IEEEmembership{Senior Member, IEEE}, \uppercase{Gheorghe Sorin Paraoanu} \authorrefmark{2,3}}

\address[1]{Department of Information and Communications Engineering, Aalto University, 02150 Espoo, Finland }
\address[2]{QTF Centre of Excellence, Department of Applied Physics,
Aalto University, 00076 Aalto, Finland}
\address[3]{InstituteQ – the Finnish Quantum Institute, Aalto University, 02150 Espoo, Finland}
\tfootnote{GSP acknowledges the project Business Finland QuTI (decision 41419/31/2020) as well as support under a research grant agreement between Saab and Aalto University. This work was performed as part of the Academy of Finland Centre of Excellence program (project 352925).}

\markboth
{Author \headeretal: Preparation of Papers for IEEE TRANSACTIONS and JOURNALS}
{Author \headeretal: Preparation of Papers for IEEE TRANSACTIONS and JOURNALS}

\corresp{Corresponding author: Hany Khalifa (e-mail: hany.khalifa@aalto.fi).}

\begin{abstract}
 Microwave quantum information networks require reliable transmission of single photon propagating modes over lossy channels. In this article we propose a microwave \textit{noise-less linear amplifier} (NLA) suitable to circumvent the losses incurred by a flying photon undergoing an \textit{amplitude damping channel} (ADC). The proposed model is constructed by engineering a simple one-dimensional four node cluster state. Contrary to conventional NLAs based on \textit{quantum scissors} (QS), single photon amplification is realized without the need for  \textit{photon number resolving detectors} (PNRDs). Entanglement between  nodes comprising the device's cluster is achieved by means of a \textit{controlled phase gate} (CPHASE). Furthermore, photon measurements are implemented by \textit{quantum non demolition detectors} (QNDs), which are currently available as a part of \textit{circuit quantum electrodynamics} (cQED) toolbox. We analyze the performance of our device practically by considering detection inefficiency and dark count probability. We further examine the potential usage of our device in low power quantum sensing applications and remote \textit{secret key generation} (SKG). Specifically, we demonstrate the device's ability to prepare loss-free resources offline, and its capacity to overcome the repeater-less bound of SKG. We compare the performance of our device against a QS-NLA for the aforementioned applications, and highlight explicitly the operating conditions under which our device can outperform a QS-NLA. The proposed device is also suitable for applications in the optical domain.
\end{abstract}


\titlepgskip=-21pt

\maketitle

\section{Introduction}
\label{sec:introduction}
\PARstart{F}{or} a multitude of bosonic microwave \textit{quantum information processing} (QIP) tasks, the reliable transmission of a single flying photon is critical to their success \cite{joshi2021quantum, cai2021bosonic}. However, due to the non-ideal transmissivity of transmission media, the accumulation of propagation losses would eventually result in a protocol failure. Seemingly, the most practical solution would be to deploy phase insensitive amplifiers in order to combat channel attenuation. Nonetheless, the process of amplification is always accompanied by at least half a quantum of noise photons \cite{caves1982quantum}. Unfortunately, a deterministic noise-free amplifier is fundamentally prohibited by virtue of the no cloning theorem \cite{wooters1982quantum}. However, a nondeterministic noiseless amplifier (NLA) is physically possible if the operation is restricted to a subset of states in the device Hilbert space \cite{ralph2009nondeterministic, pandey2013quantum}. 

Ever since the inception of NLAs, numerous experiments had been conducted in order to demonstrate the concept \cite{barbieri2011nondeterministic}. On the application level, NLAs have found great utility in fields such as  entanglement distillation \cite{xiang2010heralded, PhysRevA.100.022315}, \textit{device-independent quantum key distribution} (DI-QKD) \cite{pirandola2020advances, zapatero2023advances}, quantum repeaters \cite{xia2019repeater}, and most recently quantum teleportation \cite{zhao2023enhancing}. 

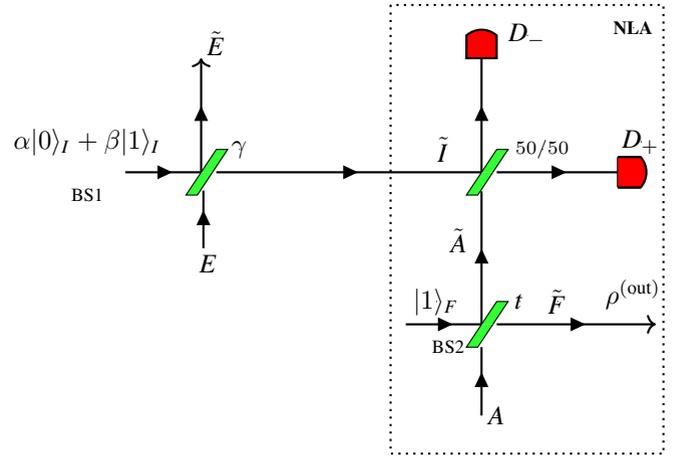
\begin{figure}
    \centering
    \begin{circuitikz}
        \draw[thick](-1.5,0) to node [currarrow, sloped] {}(-0.5,0) ;
        \draw[](-2,0.4)--(-2,0.4) node {$\alpha \lvert 0 \rangle_{I} + \beta \lvert 1\rangle_{I}$};
        \draw[fill=green!80](-0.5,0)--(-0.3,0.3)--(-0.15,0.3)--(-0.35,0);
        \draw[fill=green!80](-0.5,0)--(-0.7,-0.3)--(-0.55,-0.3)--(-0.35,0);
        \draw[](0,0.3)--(0,0.3) node[] {$\gamma$};
         \draw[](-2,-0.3)--(-2,-0.3) node[] {\scriptsize BS1 };
        \draw[thick] (-0.47,-1) to node [currarrow, sloped] {} (-0.47,-0.24);
        \draw[](-0.4,-1.2)--(-0.4,-1.2) node {$E$};
        \draw[thick](-0.3,0) to node [currarrow, sloped]{}(3.2,0);
        %\draw[thick](-0.9,0) to node [currarrow, sloped] {}(4,0) ;
        
        \draw[fill=green!80](3.2,0)--(3.4,0.3)--(3.55,0.3)--(3.35,0);
        \draw[fill=green!80](3.2,0)--(3,-0.3)--(3.15,-0.3)--(3.35,0);
        
        \draw[](4,0.3)--(4,0.3) node{\scriptsize $50/50$};
        \draw[thick](3.2,-0.25) to node [currarrow, sloped, xscale=-1] {} (3.2,-2);
        \draw[fill=green!80](3.2,-2)--(3.4,-1.7)--(3.55,-1.7)--(3.35,-2);
        \draw[fill=green!80](3.2,-2)--(3,-2.3)--(3.15,-2.3)--(3.35,-2);
        \draw[](3.7,-1.7)--(3.7,-1.7) node{$t$};
        
        \draw[](2.76,-2.3)--(2.76,-2.3) node[]{\scriptsize BS2};
        \draw[->, thick](3.4,-2) to node [currarrow, sloped]{} (5.5,-2);
        \draw[](5.2,-1.6)--(5.2,-1.6) node {$\rho^{(\text{out})}$};
        \draw[thick](2.2,-2) to node [currarrow, sloped] {} (3.2,-2);
        \draw[](2.6,-1.7)--(2.6,-1.7) node {$ \lvert 1\rangle_{F}$};
        \draw[thick](3.2,-2.3) to node [currarrow, sloped, xscale=-1] {}(3.2,-3.2);
        \draw[](3.4,-3.2)--(3.4,-3.2) node {$A$};
        \draw[thick, fill=green!70](3.40,0) to node [currarrow,sloped]{}(5,0);
        \draw[thick, fill=green!70](5,0)--(5,0.2);
        \draw[thick, fill=green!70](5,0)--(5,-0.2);
        \draw[thick, fill=green!70](5,0.2)--(5.30,0.2);
        \draw[thick, fill=green!70](5,-0.2)--(5.30,-0.2);
        \draw[thick, fill=red](5,0.2)--(5.30,0.2) to [bend left=40](5.30,-0.2)--(5,-0.2);
        \draw[thick] (3.2,0) to node [currarrow, sloped]{}(3.2,1.5);
        \draw[](3.8, 1.8)--(3.8,1.8) node {$D_{-}$};
        \draw[](5.3, 0.4)--(5.3,0.4) node {$D_{+}$};
        \draw[->,thick](-0.5,0) to node[currarrow, sloped] {} (-0.5,1.5);
        \draw[](-0.3,1.7)--(-0.3,1.7) node {$\tilde{E}$};
         \draw[](2.7,0.3)--(2.7,0.3) node {$\tilde{I}$};
        \draw[thick,fill=red] (3.2,1.5)--(3.4,1.5)--(3.4,1.8) to [bend right = 40] (3,1.8)--(3,1.5)--(3.2,1.5);   
         \draw[](4.2,-1.7)--(4.2,-1.7) node {$\tilde{F}$};
         \draw[](2.9,-0.9)--(2.9,-0.9) node {$\tilde{A}$};
         \draw[thick, dotted](2,2.2)--(2,-3.7)--(5.6,-3.7)--(5.6,2.2)--(2,2.2);
         \draw [] (5.2,1.9)--(5.2,1.9) node []{\scriptsize \textbf{NLA}};
    \end{circuitikz}
    \caption{\text{Qubit noiseless amplification}. The depicted setup is composed of three beamsplitters and two PNRDs denoted by $D_{+}$ and $D_{-}$. The first beamsplitter, BS1, models a lossy channel with transmissivity $\gamma$. The second one, BS2, has a variable transmissivity $t$, whereas the third is balanced. The output of a lossy channel is a mixture of single photon state and vacuum. The amplifier rescales the probabilities of the channel's output density operator in favor of the single photon component. Noiseless amplification  succeeds probabilistically by registering one and only one click in either $D_{+}$ or $D_{-}$.}
    %click.%      
    \label{fig:NLA}
\end{figure}

A crucial requirement for implementing NLAs as depicted in Fig. (\ref{fig:NLA}) is that the detection process that heralds a successful probabilistic amplification is accomplished by \textit{photon number resolving detectors} (PNRDs), also know as photon counters. This presents a major obstacle in front of physical implementations in the microwave domain, as the low power of single microwave photons renders the task of finding a suitable observable for photon numbers a challenging one.   

Recently the necessity of using PNRDs for the implementation of NLAs had been relaxed by proposing a new model that employs a cluster state as a resource in order to achieve noiseless amplification \cite{elemy2017one}. The concept followed the new architecture of third generation quantum repeaters \cite{azuma2015all, azuma2022quantum}, which is characterized by its fault-tolerance capabilities. In the field of quantum computation, the cluster state model is an alternative to the circuit-based framework that utilizes a resource of highly entangled qubits \cite{raussendorf2001one}. The cluster nodes (qubits) are initially prepared in the diagonal basis $\{ \lvert + \rangle, \lvert - \rangle\}$. Links (network edges) are established between nodes by means of a CPHASE gate. Every quantum computing algorithm can be realized on a cluster state resource by performing single qubit measurements on the nodes comprising the cluster in a pre-defined Pauli basis \cite{raussendorf2001one}. The same concept of cluster state computation or synonymously, the \textit{one-way model}, was later developed for the more versatile \textit{continuous variable} (CV) systems \cite{menicucci2006universal}. 

Our main objective in this article is to propose a NLA model suitable for microwave bosonic QIP. Since PNRDs are currently not available in the microwave domain, we propose utilizing  microwave QNDs, which are efficient and have low dark count rates \cite{johnson2010quantum,kono2018quantum}. Furthermore, owing to their deterministic generation process, we adopt bosonic temporal modes as an encoding scheme for all the qubits involved in our protocol \cite{da2010schemes,pechal2014microwave,brecht2015photon}. It is worth mentioning at this point that the aforementioned characteristics are essentially different from our previous proposal \cite{elemy2017one}, that relied on coincidence counters and polarization qubits. 

Our proposed model is composed of a simple 4-node one dimensional cluster state engineered in a manner that exactly simulates the dynamics of a heralded NLA as depicted in Fig. (\ref{fig:NLA}). However, as we demonstrate in the upcoming sections, this model is extremely powerful since it inherits the fault tolerance capacity of the one-way quantum computing model. 

In order to practically evaluate the performance of our proposed device, we consider two important microwave QIP tasks that could be greatly enhanced by the usage of our device. Firstly, we study the problem of storing a single bosonic mode in a leaky memory element. Then we turn our attention towards the task of entanglement generation between remote parties. We further consider the remote generation of secure keys over the established entanglement channel.

This article is organized as follows:  In Section (\Romannum{2}) we start with a formal description of the theory of NLA followed by a mathematical representation of damping and losses. Then we focus on the inner workings of a heralded NLA based on quantum scissors. Section (\Romannum{3}) is dedicated to our one-way NLA. Firstly we briefly introduce the mathematical formalism of bosonic temporal modes. After that we describe in details the working principle of our device. Then we study the effect of dark counts and non ideal efficiency on the device's success probability. Finally the last part of this section focuses on the relation between the success probability and the tuning parameter of our device. In section (\Romannum{4}) we study two important applications of our device. Firstly we tackle the problem of storing an idler mode in an entanglement-enhanced sensing protocol. We show the cost at which our device is capable of fully restoring the attenuated mode. Then we turn our attention towards another interesting problem, that is, remote entanglement generation. We first demonstrate the ability of our device to establish remote entanglement deterministically, due to its fault tolerant capacity. After that we examine the secret key rate that can be generated accordingly. We mainly focus in this final section on showing the ability of our cluster state NLA to beat the repeater-less bound of remote SKG, and we further highlight the operating conditions under which it can outperform a NLA based on QS. Finally section (\Romannum{5}) is our conclusion. 
\section{THEORY OF NLA}
 We begin this section with a brief  definition of NLA, then formally define channel losses. In the last part of this section we describe how amplification is achieved with a QS-NLA.   
\subsection{Formal definition of NLA}
Ideally a NLA can be described by defining the following operator \cite{ralph2009nondeterministic} 
\begin{align}
    \hat{T} &= g^{\hat{n}},
\end{align}
where $\lvert g \lvert  \geq 1$ is a real number characterizing the amplifier's gain, and $\hat{n}$ is the number operator. 

Consider now a coherent state $\lvert \alpha \rangle$ undergoing the aforementioned transformation
\begin{align}
    \hat{T} \lvert \alpha \rangle &= g^{\hat{n}} \Big[e^{\frac{-\lvert \alpha \lvert^{2}}{2}} \underset{n=0}{\overset{\infty}{\sum}}\frac{\alpha^{n}}{\sqrt{n!}} \lvert n\rangle \Big] \nonumber \\
    &= e^{\frac{-\lvert \alpha \lvert^{2}}{2}} \underset{n=0}{\overset{\infty}{\sum}}\frac{\alpha^{n}}{\sqrt{n!}} g^{\hat{n}} \lvert n\rangle \nonumber \\ 
    &=e^{\frac{-\lvert \alpha \lvert^{2}}{2}} \underset{n=0}{\overset{\infty}{\sum}}\frac{(g\alpha)^{n}}{\sqrt{n!}} \lvert n\rangle \nonumber \\
    &=e^{\frac{\lvert \alpha \lvert^{2}}{2} (g^{2}-1)} \lvert g \alpha \rangle,
    \label{eq:BoundNLA}
\end{align}
where $\lvert g \alpha \rangle = e^{\frac{-\lvert g\alpha \lvert^{2}}{2}} \underset{n=0}{\overset{\infty}{\sum}}\frac{(g\alpha^{n})}{\sqrt{n!}} \lvert n\rangle$, $g^{\hat{n}}=e^{\ln{g^{\hat{n}}}}=e^{(\ln{g}) \hat{n}}= \underset{n=0}{\overset{\infty}{\sum}}[(\ln{g}) \hat{n}]^{n}/ n!$, $\hat{n} \lvert n \rangle = n \lvert n \rangle$, and $P_{d} = e^{\lvert \alpha \lvert^{2}(g^{2}-1)}$ is the success probability of this transformation which equals to the norm of the output state \cite{ralph2009nondeterministic}. 

A linear operator in a Hilbert space, $\hat{O} \in \mathcal{L}(\mathcal{H})$, $\hat{O} : \mathcal{H}_{1} \rightarrow \mathcal{H}_{2}$, is bounded if there exists a constant $\mathcal{K} \leq 0$ such that \cite{christensen2010functions}
\begin{align}
    \lvert \lvert \hat{O} v \lvert \lvert_{\mathcal{H}_{2}} \leq \mathcal{K} \lvert \lvert v \lvert \lvert_{\mathcal{H}_{1}} \hspace{0.5em} \forall v \in \mathcal{H}_{1}
\end{align}
where $\mathcal{H}_{1}$, $\mathcal{H}_{2}$ are the domain and target Hilbert spaces respectively, $v$ is a ket vector, and the smallest possible value of $\mathcal{K}$ is called the operator norm, which is denoted by $\lvert \lvert \hat{O} \lvert \lvert$.

The transformation described in Eq. (\ref{eq:BoundNLA}) effectively multiplies the coherent field input amplitude $\alpha$ by a gain factor $g$ without the addition of any extra noise, $\hat{T}\lvert \alpha \rangle \rightarrow  c \lvert g \alpha \rangle$, where $c$ is a complex number satisfying $0<\lvert c \lvert<1$. However, in order for this transformation to be valid for all gain values, the operator $\hat{T}$ has to be unbounded, and hence the entire process becomes nonphysical. Overcoming this problem was achieved in the original proposal \cite{ralph2009nondeterministic} by splitting the input coherent state evenly on an $N$-port beamsplitter, such that each individual photon comprising the original input enters a quantum scissor (QS) device \cite{PhysRevLett.81.1604} similar to the one depicted in Fig. (\ref{fig:NLA}). The overall success of the device is the independent successes of each QS, which is heralded by registering one and only photon in either $D_{+}$ or $D_{-}$. Finally the outputs of the $N$ quantum scissors are interferometrically recombined on a second N-port beamsplitter in order to produce the desired amplified output. When $N$ is large, the success probability of this device becomes $P_{d} \approx (1-t)^{\frac{N}{2}} e^{\lvert \alpha \lvert^{2}(g^{2}-1)}$ which decreases when the variable transmissivity (see Fig. (\ref{fig:NLA})) $t>1/2$, and $N$ is large. Consequently this bounds the operator's norm which deems the transformation physical.

The vanishing success probability as the number of ports increases implies that NLA is physically tangible only when considering a truncation of the original input state. Alternatively by limiting the device's operation to small input states, $\lvert \alpha \lvert \ll 1$, the operator $\hat{T}$ can be made physical by using only one QS.  
\subsection{Channel model}
In bosonic QIP information loss is dominantly due to energy dissipative processes. Mathematically a lossy transmission medium is described as an \textit{amplitude damping channel} (ADC) \cite{nielsen2010quantum}. In the language of quantum operations, ADC transforms an input \textit{positive semidefinite} (PSD) density operator $\rho_{\text{in}} \geq 0$, into an output PSD, thus rendering it a \textit{completely positive trace preserving map} (CPTP) \cite{stinespring1955positive}. The quantum operation corresponding to a qubit ADC can be written in terms of its Kraus representation \cite{kraus1983states}, $\mathcal{E}_{\text{AD}}(\rho_{\text{in}}) = \underset{k=0}{\overset{1}{\sum}}K^{}_{k}\rho_{\text{in}} K^{\dagger}_{k}=K^{}_{0} \rho_{\text{in}}K^{\dagger}_{0}+ K^{}_{1}\rho_{\text{in}}K^{\dagger}_{1}$, where $K_{0} = \left(\begin{smallmatrix}
    1 &0 \\
    0&\sqrt{\gamma}
\end{smallmatrix}\right)$, $K_{1} = \left(\begin{smallmatrix}
    0 & \sqrt{1-\gamma} \\
    0&0
\end{smallmatrix}\right)$ satisfy $\underset{k=0}{\overset{1}{\sum}} K^{\dagger}_{k}K^{}_{k} = I$, such that $I$ is the two dimensional identity matrix, and $\gamma $ is the survaivability of one photon. Practically, a qubit ADC can be modelled by a unitary passive beamsplitter transformation, where the beamsplitter's unused port injects a vacuum mode, $U_{\theta} = e^{-i \theta H/\hbar}$, where $H=i(a^{\dagger}b+b^{\dagger}a)$ is the beamsplitter's Hamiltonian, $\theta$ is its angle, and $a$ and $b$ are its first and second modes respectively. An arbitrary single photon state state occupying the first mode evolves under the previous Hamiltonian as $U_{\theta } (\alpha\lvert 0_{a}\rangle + \beta \lvert 1_{a} \rangle) \otimes \lvert0_{b} \rangle = (\alpha \lvert 0_{a} \rangle+\beta \cos{\theta}\lvert 1_{a} \rangle)\otimes \lvert 0_{b}\rangle+\beta \sin{\theta}\lvert 0_{a}\rangle \otimes \lvert 1_{b}\rangle$, where $ U^{}_{\theta }a^{\dagger}U^{\dagger}_{\theta }U^{}_{\theta } \lvert 0_{a}0_{b} \rangle =U^{}_{\theta }a^{\dagger}U^{\dagger}_{\theta }\lvert 0_{a}0_{b} \rangle = a^{\dagger} \cos{\theta}+b^{\dagger} \sin{\theta} \lvert 0_{a}0_{b} \rangle$. Then, tracing out the environment mode $b$ leaves the input state in a mixture, $\rho_{\text{out}} =  (1/(1+\beta^{2}\sin^{2}{\theta}))\lvert \Psi \rangle \langle \Psi \lvert +  \big((\beta^{2}\sin^{2}{\theta})/(1+\beta^{2}\sin^{2}{\theta})\big) \lvert 0 \rangle \langle 0 \lvert$, where $\lvert \Psi \rangle = (\alpha \lvert 0 \rangle + \beta \cos{\theta} \lvert 1 \rangle)/(\sqrt{\alpha^{2}+\beta^{2} \cos^{2}{\theta}})$. The Krauss representation defined earlier can yield the same output mixed state by defining $\gamma = \cos^{2}{\theta}$, and thus $1-\gamma = \sin^{2}{\theta}$ would be the probability that a photon is lost. 

It is useful at this point to introduce the gate-based model of amplitude damping. This will be recalled when we develop our one-way NLA. Consider an arbitrary single photon state, $\alpha \lvert 0 \rangle_{c} + \beta \lvert 1 \rangle_{c} $, as the control of a two-qubit controlled rotation gate, $\lvert 0 \rangle_{c} \langle 0 \lvert_{c} \otimes I_{t} + \lvert 1\rangle_{c} \langle 1 \lvert_{c} \otimes R^{\theta}_{t}$, where $R^{\theta} = \left(\begin{smallmatrix}
   \cos{\theta} &-\sin{\theta} \\
    \sin{\theta}&\cos{\theta}
\end{smallmatrix}\right)$, and the gate's target is prepared in a vacuum state $\lvert 0 \rangle_{t}$. The gate's output can be written as, $\big[\alpha \lvert 0\rangle_{c} \otimes \lvert 0\rangle_{t}\big]+ \big[\beta \lvert 1\rangle_{c} \otimes \big(\cos{\theta} \lvert 0 \rangle_{t}+\sin{\theta} \lvert 1 \rangle_{t}\big)\big]$. Then using the same output as an input to a \textit{controlled not} (CNOT) entangling gate, $\text{CNOT} = \lvert 0 \rangle_{c} \langle 0 \lvert_{c} \otimes I_{t} + \lvert 1 \rangle_{c} \langle 1 \lvert_{c} \otimes X_{t} $, where $X$ is the Pauli bit flip operation, and exchanging the roles of control and target qubits, yields the following  output state, $\alpha \lvert 0_{t}\rangle \otimes \lvert 0_{c}\rangle+ \beta\cos{\theta} \lvert 1_{t}\rangle \otimes \lvert 0_{c} \rangle+\beta \sin{\theta} \lvert 0_{t} \rangle \lvert 1_{c} \rangle$. Finally after tracing out the control qubit, it can be straightforwardly verified that the output is similar to that of the aforementioned ADC.   

An equivalent model for a qubit ADC can be obtained by using a controlled phase operation (CPHASE), two Hadamard gates and employing the relation $X=HZH$, where $H$ is a Hadamard gate, and $Z$ is a Pauli phase flip operator \cite{nielsen2010quantum}. This will be more clear in the next section when developing our device model. 

\subsection{ QS-NLA}
Fig. (\ref{fig:NLA}) depicts an experimental setup of a NLA based on a single QS. At a transmitter an arbitrary single photon state  $\alpha \lvert 0\rangle_{I}  + \beta \lvert 1 \rangle_{I}$, where $\lvert \alpha \lvert^{2}+ \lvert \beta \lvert^{2}=1$, is prepared. The photon undergoes an ADC characterized by a unitary beamsplitter transformation as described in the previous section. By choosing an appropriate beamsplitter angle $\theta$, an ADC can be simply written as a $2\times2$ matrix, ${\rm BS1}=\left( \begin{smallmatrix}
    \sqrt{\gamma} & -\sqrt{1-\gamma} \\ 
    \sqrt{1-\gamma} & \sqrt{\gamma}
\end{smallmatrix}\right)$. In the Heisenberg picture the output modes are transformed by the beamsplitter matrix as 
\begin{align}
    \tilde{I} &= \sqrt{\gamma} I + \sqrt{1-\gamma} E \nonumber \\ 
    \tilde{E} &= \sqrt{\gamma} E - \sqrt{1-\gamma} I, 
\end{align}
whereas, in the Schr\"{o}dinger picture an input state vector evolves as  %Correspondingly the overall state state leaving the channel is written as% 
\begin{align}
    \lvert \Phi\rangle_{IE}  & = (\alpha \lvert 0 \rangle_{I} + \beta \lvert 1 \rangle_{I})\otimes \lvert 0 \rangle_{E} \nonumber \\ 
    &= \alpha \lvert 0 0 \rangle_{{I}{E}} + \beta  I^{\dagger}  \lvert 0 0 \rangle_{{I}{E}} \nonumber \\
    \lvert \Phi \rangle_{\tilde{I}\tilde{E}}  & = (\alpha \lvert 0  \rangle_{\tilde{I}} + \sqrt{\gamma} \beta \lvert 1 \rangle_{\tilde{I}}) \otimes \lvert 0 \rangle_{\tilde{E}} - \sqrt{1-\gamma} \beta \lvert 0 \rangle_{\tilde{I}} \otimes \lvert 1\rangle_{\tilde{E}}. 
    \label{eq:ChannelLoss}
\end{align}
%At the receiving end%

Simultaneously an auxiliary single photon is mixed with a vacuum input on a second beamsplitter defined by ${\rm BS2} =\left( \begin{smallmatrix}
    \sqrt{t} & -\sqrt{1-t} \\ 
    \sqrt{1-t} & \sqrt{t}
\end{smallmatrix}\right)$. We assume that the beamsplitter's transmissivity $t$ is controllable and can be adjusted as desired. Accordingly the beamsplitter's modes transform as %Consequently, the output modes become%
\begin{align}
    \tilde{F} &= \sqrt{t} F - \sqrt{1-t} A \nonumber \\ 
    \tilde{A} &= \sqrt{t} A + \sqrt{1-t} F,  
\end{align}
and hence the auxiliary photon state vector evolves as 
\begin{align}
    F^{\dagger}\lvert 0 \rangle_{A} \lvert 0 \rangle_{F} &= (  \sqrt{t} \tilde{F}^{\dagger}+\sqrt{1-t} \tilde{A}^{\dagger}) \lvert 0 0 \rangle_{\tilde{A}\tilde{F}} \nonumber \\ 
    \lvert \Phi \rangle_{\tilde{A}\tilde{F}} &= \sqrt{1-t} \lvert 10 \rangle_{\tilde{A}\tilde{F}}  + \sqrt{t} \lvert 0 1 \rangle_{\tilde{A}\tilde{F}}.
    \label{eq:Auxresource}
\end{align}
The two modes, $\Tilde{I}$, $\Tilde{A}$ are then redirected towards a balanced beamsplitter for a \textit{Bell state measurement} (BSM). They evolve according to
\begin{align}
    \Tilde{I} &= \frac{1}{\sqrt{2}}(D_{+}-D_{-}) \nonumber \\ 
    \Tilde{A} &= \frac{1}{\sqrt{2}}(D_{+}+D_{-}),
    \label{eq:clicks}
\end{align}
where $D_{+}$ and $D_{-}$ are detector operators. 

Hence the overall state becomes
\begin{align}
    \lvert \Phi^{(\mathrm{out})} \rangle &= \lvert \Phi \rangle_{\tilde{I}\tilde{E}} \otimes \lvert \Phi \rangle_{\tilde{A}\tilde{F}} \nonumber \\ 
    &= (\alpha \lvert 0 0\rangle_{\tilde{I}\tilde{E}} + \beta\sqrt{\gamma}\lvert 10\rangle_{\tilde{I}\tilde{E}} -\beta \sqrt{1-\gamma} \lvert 0 1 \rangle_{\tilde{I}\tilde{F}}) \nonumber \\
    & \otimes(\sqrt{1-t} \lvert 1 0 \rangle_{\tilde{A}\tilde{F}} + \sqrt{t} \lvert 0 1 \rangle_{\tilde{A}\tilde{F}}) \nonumber \\ 
    &= \alpha \sqrt{1-t} \lvert 0010 \rangle_{\tilde{I}\tilde{E}\tilde{A}\tilde{F}}+ \alpha \sqrt{t} \lvert 0001 \rangle_{\tilde{I}\tilde{E}\tilde{A}\tilde{F}} \nonumber \\ 
    &+ \beta \sqrt{\gamma(1-t)}\lvert 1010 \rangle_{\tilde{I}\tilde{E}\tilde{A}\tilde{F}} +\beta \sqrt{\gamma t }  \lvert 1001 \rangle_{\tilde{I}\tilde{E}\tilde{A}\tilde{F}} \nonumber \\ 
    &-\beta\sqrt{(1-\gamma)(1-t)}  \lvert 0110 \rangle_{\tilde{I}\tilde{E}\tilde{A}\tilde{F}} \nonumber \\ &- \beta\sqrt{(1-\gamma)t} \lvert 0101 \rangle_{\tilde{I}\tilde{E}\tilde{A}\tilde{F}} \nonumber \\ 
    &= (\alpha \sqrt{1-t} \Tilde{A}^{\dagger}+\alpha \sqrt{t}\tilde{F}^{\dagger}+ \beta \sqrt{\gamma (1-t)}\tilde{I}^{\dagger}\tilde{A}^{\dagger}\nonumber \\ 
    &+\beta \sqrt{\gamma t}\tilde{I}^{\dagger}\tilde{F}^{\dagger}- \beta \sqrt{(1-\gamma)(1-t)}\tilde{E}^{\dagger}\tilde{A}^{\dagger}\nonumber \\ 
    &-\beta \sqrt{(1-\gamma)t}\tilde{E}^{\dagger}\tilde{F}^{\dagger})\lvert 0000 \rangle_{\tilde{I}\tilde{E}\tilde{A}\tilde{F}}.
    \label{eq:overall}
\end{align}

The required event for a successful amplification occurs when one and only one detector clicks. Thus, when $D^{+}$ clicks, a BSM projects the overall state to 
\begin{align}
    \lvert \Phi^{(\rm{out})}_{+} \rangle &= \frac{1}{\sqrt{2}}(\alpha \sqrt{1-t} \lvert 0\rangle_{\tilde{F}} + \beta \sqrt{\gamma t} \lvert 1 \rangle_{\tilde{F}}) \otimes \lvert 0 \rangle_{\tilde{E}} \nonumber \\ 
    &-\frac{1}{\sqrt{2}}(\beta \sqrt{(1-\gamma)(1-t)}) \lvert 0\rangle_{\tilde{F}} \otimes \lvert1 \rangle_{\tilde{E}},
\end{align}
whereas when $D^{-}$ clicks the overall state becomes
\begin{align}
    \lvert \Phi^{(\rm{out})}_{-}  \rangle &= \frac{1}{\sqrt{2}}(\alpha \sqrt{1-t} \lvert 0\rangle_{\tilde{F}} - \beta \sqrt{\gamma t} \lvert 1 \rangle_{\tilde{F}}) \otimes \lvert 0 \rangle_{\tilde{E}} \nonumber \\ 
    &-\frac{1}{\sqrt{2}}(\beta \sqrt{(1-\gamma)(1-t)}) \lvert 0\rangle_{\tilde{F}} \otimes \lvert1 \rangle_{\tilde{E}},
\end{align}
where the single photon sign flip in the previous equation can be fixed by an application of a Pauli phase flip operator $Z$ on mode $\tilde{F}$.

Tracing out the environment mode $\tilde{E}$ yields the density operator of the amplifier's output 
\begin{align}
    \rho^{\text{(out)}} &= \frac{1}{N} \big[  \lvert \Psi^{\text{(out)}} \rangle \langle \Psi^{\text{(out)}} \lvert + \beta^{2} (1-\gamma)(1-t) \lvert 0 \rangle \langle 0 \lvert  \big],
\end{align}
where $\lvert \Psi^{\text{(out)}} \rangle = \alpha \sqrt{(1-t)} \lvert 0\rangle_{\tilde{F}} + \beta \sqrt{\gamma t} \lvert 1 \rangle_{\tilde{F}}$, and $N= \alpha^{2}(1-t)+ \beta^{2} [\gamma t + (1-\gamma)(1-t)]$ is a normalization constant.

The transformation applied by a NLA on the input state lead to a change in the weightings of the vacuum and single photon components. As can be seen, an increase in the value of the tunable parameter $t$ rescales the input photon probability amplitudes in favor of the single photon component. The purely vaccuum output corresponds to an error event or more precisely a failed amplification attempt which embodies the probabilistic nature of such device.  

The success probability of the amplifier can be calculated as 
\begin{align*}
    P_{\mathrm{succ}} &= 2 \langle \Phi^{(\rm{out})}_{+}  \lvert \Phi^{(\rm{out})}_{+}  \rangle \nonumber \\
    &= \alpha^{2}(1-t)+\beta^{2}[(\gamma t + (1-\gamma)(1-t)],
\end{align*}
where the factor of $2$ comes from the fact that a click in either $D_{+}$ or $D_{-}$ signifies a successful amplification.

The implementation of the previous amplification protocol in the microwave domain is extremely challenging, since microwave single photon counters are still at the conceptual level. Furthermore, when a QS-NLA fails, that is, none of the detectors register any photon or they capture both photons, the entire process has to be repeated again, since the amplifier's output in this case is two uncorrelated error modes. By exploiting a cluster state of entangled photons, we demonstrate in the next section that we can achieve microwave NLA by using the available microwave technology. Furthermore, we show that our device is fault-tolerant and can operate on the protocol's failed instances to correct an error state, which is a feature that cannot be attained by QS-NLA. 
\section{MICROWAVE ONE-WAY NLA}
For microwave bosonic QIP, the most efficient state of the art encoding scheme is photon temporal modes (TMs) \cite{brecht2015photon, pechal2014microwave}. This is due to the unprecedented  controllability of light-matter couplings in cQED systems that facilitated the generation and absorption of single photon wavepackets. Thus we spend the first part of this section introducing the mathematical formalism of bosonic temporal modes \cite{brecht2015photon}. After that we proceed with explaining the inner workings of our device.
\subsection{Bosonic temporal modes}
A single bosonic mode occupying a particular TM can be represented as a coherent superposition of a continuum of single bosonic monochromatic modes 
\begin{align}
    \ket{1_{\alpha}} = \underset{- \infty}{\overset{\infty}{\int}} d  \omega \hspace*{0.2em} \alpha_{j}(\omega) a^{\dagger}(\omega) \ket{0},
\end{align}
where $\alpha_{j}(\omega)$ is a complex-valued spectral function of the corresponding wave-packet, $\underset{- \infty}{\overset{\infty}{\int}} d \omega \lvert \alpha(\omega)\rvert ^{2} = 2\pi$, $a^{\dagger}(\omega)$ is a monochromatic creation operator and the vacuum state is assumed to be multimode, i.e., $\underset{m=1}{\overset{\infty}{\bigotimes}} \ket{0}$.

In the time domain the previous representation can be interpreted as a coherent superposition of a continuum of \textit{creation times} 
\begin{align}
    \ket{1_{\alpha}} &= \underset{- \infty}{\overset{\infty}{\int}} dt \hspace*{0.2em} \alpha_{j}(t) a^{\dagger}(t) \ket{0} \nonumber \\
    &= \tilde{a}_{j}^{\dagger}\ket{0},
\end{align}
where $\tilde{a}_{j}^{\dagger}$ is denoted as broadband operator,  and the following Fourier relations were assumed 
\begin{align}
    a(t) &= \frac{1}{2 \pi} \underset{- \infty}{\overset{\infty}{\int}} d \omega e^{-i \omega t} a(\omega) \nonumber\\
      a(\omega) &= \underset{- \infty}{\overset{\infty}{\int}} dt e^{i \omega t} a(t) \nonumber \\
       \alpha(t) &=  \underset{- \infty}{\overset{\infty}{\int}} d \omega e^{-i \omega t} \alpha(\omega)\nonumber \\
        \alpha(\omega) &= \frac{1}{2\pi} \underset{- \infty}{\overset{\infty}{\int}} d \omega e^{i \omega t} \alpha(t).
\end{align}

Together with the convention that the adjoint operator is defined as $a^{\dagger}(\omega) = [a(-\omega)]^{\dagger} = \underset{- \infty}{\overset{\infty}{\int}} d\omega e^{i \omega t}a(t)$, and the wave-packet bosonic commutation relations $[a(\omega), a^{\dagger}(\omega')]= 2\pi \delta(\omega-\omega')$, whereas in the time-domain $[a(t), a^{\dagger}(t')]=\delta(t-t')$, we have a complete characterization of the creation and annihilation operators in each of the frequency and time domains.\\
TMs comprise a basis for a single boson wavepacket. Thus any arbitrary single bosonic mode can be expanded as a linear superposition of TMs 
\begin{align}
    \ket{\Psi} = \underset{j=0}{\overset{\infty}{\sum}} c_{j} \tilde{a}_{j}^{\dagger} \ket{0},
\end{align}
where $c_{j}$ is a complex amplitude.

Accordingly, a 2-dimensional bosonic space can be defined, such that any arbitrary qubit is written as 
\begin{align}
    \ket{\Psi} = \underset{j=0}{\overset{1}{\sum}} c_{j} \tilde{a}_{j}^{\dagger} \ket{0},
    \label{eq:qubitSpace}
\end{align}
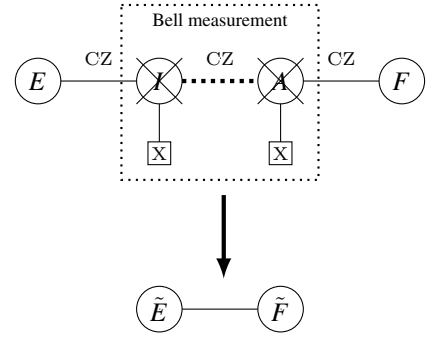
\begin{figure}
    \centering
    \label{fig:ONEWAYNLA}
\begin{tikzpicture}
   \filldraw[fill=white] (0,0) circle (0.3); 
   \draw[](0.3,0)--(1.3,0);
   \filldraw[fill=white] (1.6,0) circle (0.3); 
   \filldraw[fill=white] (3.2,0) circle (0.3); 
   \draw[](3.5,0)--(4.5,0);
    \filldraw[fill=white] (4.8,0) circle (0.3); 
    \draw [ultra thick, dotted] (1.9,0)--(2.9,0);
    \draw[](1.6,-0.3)--(1.6,-0.8);
     \draw[](3.2,-0.3)--(3.2,-0.8);
     \draw[](1.6,-0.8)--(1.75,-0.8)--(1.75,-1.1)--(1.45,-1.1)--(1.45,-0.8)--(1.6,-0.8);
       \draw[](3.2,-0.8)--(3.35,-0.8)--(3.35,-1.1)--(3.05,-1.1)--(3.05,-0.8)--(3.2,-0.8);
       \draw[](1.6,-0.95)--(1.6,-0.95) node [] {\scriptsize$\mathrm{X}$};
       \draw[](3.2,-0.95)--(3.2,-0.95) node [] {\scriptsize$\mathrm{X}$};
    \draw [] (0,0)--(0,0) node [] { $E$ };
    \draw [] (1.6,0)--(1.6,0) node [] { $I$ };
       \draw [] (3.2,0)--(3.2,0) node [] { $A$ };
       \draw [] (4.8,0)--(4.8,0) node [] { $F$ };
       \draw [] (0.8,0.3)--(0.8,0.3) node [] {\scriptsize $\mathrm{CZ}$ };
       \draw [] (2.4,0.3)--(2.4,0.3) node [] {\scriptsize $\mathrm{CZ}$ };
       \draw [] (4,0.3)--(4,0.3) node [] {\scriptsize $\mathrm{CZ}$ };
       \draw [] (1.6,0) node[cross,black] {};
       \draw [] (3.2,0) node[cross,black] {};
       \draw[-latex, ultra thick] (2.45,-1.5)--(2.45,-2.6);
       \draw[thick, dotted] (1.1,1)--(3.7,1)--(3.7,-1.3)--(1.1,-1.3)--(1.1,1);
       \draw[] (2.4,0.8)--(2.4,0.8) node []{\scriptsize Bell measurement};
        \filldraw[fill=white] (1.6,-3) circle (0.3);
       \filldraw[fill=white] (3.2,-3) circle (0.3);
       \draw [] (1.9,-3)--(2.9,-3);
        \draw [] (1.6,-3)--(1.6,-3) node [] { $\tilde{E}$ };
        \draw [] (3.2,-3)--(3.2,-3) node [] { $\tilde{F}$ };
\end{tikzpicture}
\caption{A schematic of the one-way NLA. The amplification process is composed of three steps. First \textit{state preparation}, this is done by entangling each of the  $I-E$ and $A-F$ pairs by means of a CZ operation. After that, a BSM is performed on the middle nodes, $I$ \& $A$. This is achieved by first entangling the two nodes via a CZ gate followed by two projective X-basis measurements. As a result, the two nodes are removed from the cluster and  nodes $E$ \& $F$ are directly entangled. Finally, according to the measurement outcomes we  either perform  error correction or not. The protocol is a reminiscent of an entanglement swapping based quantum repeater }
\label{fig:ONEAMP}
\end{figure}
where the logical zero basis state is defined as, $\ket{\tilde{0}} = \tilde{a}^{\dagger}_{0}\lvert 0 \rangle = \underset{- \infty}{\overset{\infty}{\int}} d  \omega \hspace*{0.2em} \alpha_{j}(\omega) a^{\dagger}(\omega) \ket{0}$, whereas the logical one is, $\ket{\tilde{1}} = \tilde{a}^{\dagger}_{1} \lvert 0 \rangle =\underset{- \infty}{\overset{\infty}{\int}} d  \omega \hspace*{0.2em} \beta_{j}(\omega) a^{\dagger}(\omega) \ket{0}$, such that $\langle \alpha_{j} \lvert \beta_{j} \rangle = \underset{-\infty}{\overset{\infty}{\int}} d \omega \Bar{\alpha_{j}} \beta_{j} = 0$ are orthonormal complex temporal modes. 

Experimentally an arbitrary TM qubit is prepared by splitting the photon's wavepacket on a beamsplitter, more precisely its microwave counterpart, the quadrature hybrid \cite{pozar2011microwave, da2010schemes, hoffmann2010superconducting}, $\tilde{a}^{\dagger}_{b} \lvert 00\rangle= (\sqrt{T}\tilde{a}^{\dagger}_{1}+\sqrt{1-T}\Tilde{a}^{\dagger}_{v}) \lvert 00\rangle = c_{0} \lvert \tilde{0}\rangle + c_{1} \lvert \tilde{1}\rangle$,  where $\tilde{a}_{b}$ is the beamsplitter's output, $T$ is its transmsissivity, $c_{0}$, $c_{1}$ are equal to $\sqrt{1-T}$, $\sqrt{T}$ respectively and we have assumed a dual-rail encoding defined as $\lvert 0_{1}1_{v} \rangle \rightarrow\lvert \Tilde{0}\rangle$, $\lvert 1_{1}0_{v} \rangle \rightarrow \lvert \tilde{1}\rangle$. It is also worth mentioning that a Hadamard transformation can be realized in a similar way \cite{gerry2005introductory}.

The qubit space defined in Eq. (\ref{eq:qubitSpace}) can be regarded as analogous to the two dimensional polarization space, where the basis vectors $\{\lvert H \rangle, \lvert V \rangle \}$ correspond to the bosonic polarization state. For ease of readability we drop all tildes, subscripts from all qubit TMs and their corresponding operators, since it is now clear that a qubit's TM space is \textit{isomorphic} \cite{nielsen2010quantum} to a conventional two-dimensional Hilbert space.   
\subsection{Device description}
The core idea of our device is that a cluster state model is capable of simulating the dynamics of any circuit-based quantum algorithm. Furthermore, the model's inherent redundancy guarantees fault tolerance. Our task in this section is to show that the setup described in Fig. (\ref{fig:NLA}) can be simulated by a cluster model (see Fig. (\ref{fig:ONEAMP})) with the capacity of being immune to failed attempts, thus promising an uninterrupted performance. 

Consider an arbitrary single photon undergoing a passive loss channel, an \textit{amplitude damping channel} (ADC), as previously described. By adopting the gate-based ADC model explained earlier, the channel's dynamics can be described as follows 
\begin{align}
   \lvert \Phi \rangle_{EI} &= (\alpha \lvert 0 \rangle_{E}+\beta \lvert 1 \rangle_{E}) \otimes \lvert 0 \rangle_{I} \nonumber \\ 
   {\rm{CR}}^{\theta} \lvert \Phi \rangle_{EI} &= \alpha \lvert 00 \rangle_{EI}+\beta \sqrt{1-\gamma} \lvert 10 \rangle_{EI} \nonumber \\
   &+ \beta \sqrt{\gamma} \lvert 11 \rangle_{EI}, \nonumber \\ 
   ({\rm{H}} \otimes {\rm{I}}){\rm{CR}}^{\theta} \lvert \Phi \rangle_{EI} &= \alpha \lvert +0 \rangle_{EI}+\beta \sqrt{1-\gamma} \lvert -0 \rangle_{EI} \nonumber \\ 
   &+ \beta \sqrt{\gamma} \lvert -1 \rangle_{EI} \nonumber \\ 
   {\rm{CZ}}({\rm{H}} \otimes {\rm{I}}){\rm{CR}}^{\theta} \lvert \Phi \rangle_{EI} &=\alpha \lvert +0 \rangle_{EI}+\beta \sqrt{1-\gamma} \lvert -0 \rangle_{EI} \nonumber \\ 
   &+ \beta \sqrt{\gamma} \lvert +1 \rangle_{EI}, \nonumber 
   \end{align}
 finally,
 \begin{align}
   &({\rm{H}} \otimes {\rm{I}}){\rm{CZ}}({\rm{H}} \otimes {\rm{I}}){\rm{CR}}^{\theta} \lvert \Phi \rangle_{EI} =  \lvert \Phi \rangle_{\tilde{E}\tilde{I}},  \nonumber \\ 
    &\lvert \Phi \rangle_{\tilde{I}\tilde{E}} =(\alpha \lvert 0 \rangle_{\tilde{I}}+\beta \sqrt{\gamma} \lvert 1 \rangle_{\tilde{I}}) \otimes \lvert 0 \rangle_{\tilde{E}} + \beta \sqrt{1-\gamma} \lvert 0 \rangle_{\tilde{I}} \otimes \lvert 1 \rangle_{\tilde{E}},
 \end{align}
where $\mathrm{CZ} = \lvert 0 \rangle_{c} \langle 0 \lvert_{c} \otimes \mathrm{I}_{t}+\lvert 1 \rangle_{c} \langle 1 \lvert_{c} \otimes \mathrm{Z}_{t}$, $\mathrm{I} = \lvert 0 \rangle \langle 0 \lvert+\lvert 1 \rangle \langle 1 \lvert$, $\mathrm{Z} = \lvert 0 \rangle \langle 0 \lvert - \lvert 1 \rangle \langle 1 \lvert$, $\mathrm{H} = \frac{1}{\sqrt{2}}(\lvert 0 \rangle \langle 0 \lvert+\lvert 0 \rangle \langle 1 \lvert+\lvert 1 \rangle \langle 0 \lvert-\lvert 1 \rangle \langle 1 \lvert)$, and $\lvert 0 \rangle_{c} \langle 0 \lvert_{c} \otimes {\rm{I}}_{t} + \lvert 1\rangle_{c} \langle 1 \lvert_{c} \otimes {\rm{R}}^{\theta}_{t}$, where ${\rm{R} }^{\theta} = \cos{\theta}\lvert 0 \rangle \langle 0 \lvert$ $-\sin{\theta} \lvert 0 \rangle \langle 1 \lvert + \sin{\theta } \lvert 1 \rangle \langle 0 \lvert +\cos{\theta} \lvert 1 \rangle \langle 1 \lvert$, $\lvert \alpha \lvert ^{2}+ \lvert \beta \lvert^{2}=1$, and $\gamma = \cos^{2}{\theta}$. 

Simultaneously two resource modes are prepared, such that, one is prepared as a $\lvert + \rangle_{F} = (1/\sqrt{2})(\lvert 0 \rangle_{F}+\lvert 1\rangle_{F})$, while the other is a weighted superposition $\lvert \Phi \rangle_{A}=\sqrt{1-t} \lvert 0 \rangle_{A} + \sqrt{t} \lvert 1 \rangle_{A}$, where $t$ serves as a tunable parameter as before. Then a CZ entangles the two modes in a simple 2-node cluster \cite{browne2005resource} as follows 
\begin{align}
    &\lvert \Phi \rangle_{AF} = \lvert \Phi \rangle_{A} \otimes \lvert + \rangle_{F} \nonumber \\ 
    &\lvert \Phi \rangle_{AF}= \sqrt{\frac{1-t}{2}} \lvert 0 0\rangle_{AF} + \sqrt{\frac{1-t}{2}} \lvert 01 \rangle_{AF}+\sqrt{\frac{t}{2}} \lvert 10\rangle_{AF} \nonumber \\ &+\sqrt{\frac{t}{2}} \lvert 11 \rangle_{AF} \nonumber \\   
    &\mathrm{CZ} \lvert \Phi \rangle_{AF}= \lvert \Phi \rangle_{\tilde{A}\tilde{F}} \nonumber \\
    &\lvert \Phi \rangle_{\tilde{A}\tilde{F}}= \sqrt{\frac{1-t}{2}} \lvert 0 0\rangle_{\tilde{A}\tilde{F}} + \sqrt{\frac{1-t}{2}} \lvert 01 \rangle_{\tilde{A}\tilde{F}}+\sqrt{\frac{t}{2}} \lvert 10\rangle_{\tilde{A}\tilde{F}} \nonumber \\ &-\sqrt{\frac{t}{2}} \lvert 11 \rangle_{\tilde{A}\tilde{F}} \nonumber,
    \end{align}
then,
    \begin{align}
   &\lvert \Phi \rangle_{\tilde{A}\tilde{F}} = \sqrt{1-t} \lvert 0+\rangle_{\tilde{A}\tilde{F}} + \sqrt{t} \lvert 1-\rangle_{\tilde{A}\tilde{F}} \nonumber \\ 
   &( \mathrm {I}\otimes \mathrm {H}) \lvert \Phi \rangle_{\tilde{A}\tilde{F}} = \sqrt{1-t} \lvert 00\rangle_{\tilde{A}\tilde{F}} + \sqrt{t} \lvert 11\rangle_{\tilde{A}\tilde{F}} 
\end{align}
After that modes $\tilde{I}$, and $\tilde{A}$ enter another CZ gate, such that mode $\tilde{I}$ is the gate's control, whereas mode $\tilde{A}$ is its target 
\begin{align}
    &\lvert \Phi^{({\rm{out}})} \rangle \nonumber \\ 
    &=  \lvert \Phi \rangle_{\tilde{I}\tilde{E}} \otimes \lvert \Phi \rangle_{\tilde{A}\tilde{F}} \nonumber \\
    &= \alpha\sqrt{1-t} \lvert 0000\rangle_{\tilde{I}\tilde{E}\tilde{A}\tilde{F}}+\alpha\sqrt{t}\lvert 0011\rangle_{\tilde{I}\tilde{E}\tilde{A}\tilde{F}} \nonumber \\ &+\beta \sqrt{\gamma(1-t)} \lvert 1000 \rangle_{\tilde{I}\tilde{E}\tilde{A}\tilde{F}}+ \beta \sqrt{\gamma t} \lvert 1011 \rangle_{\tilde{I}\tilde{E}\tilde{A}\tilde{F}} \nonumber \\ & +\beta \sqrt{(1-\gamma)(1-t)} \lvert 0100 \rangle_{\tilde{I}\tilde{E}\tilde{A}\tilde{F}}+\beta\sqrt{t(1-\gamma)} \lvert 0111 \rangle_{\tilde{I}\tilde{E}\tilde{A}\tilde{F}}, \nonumber \\ 
    &\mathrm{CZ}  \lvert \Phi^{({\rm{out}})} \rangle \nonumber \\
    &= \alpha\sqrt{1-t} \lvert 0000\rangle_{\tilde{I}\tilde{E}\tilde{A}\tilde{F}}+\alpha\sqrt{t}\lvert 0011\rangle_{\tilde{I}\tilde{E}\tilde{A}\tilde{F}} \nonumber \\ &+\beta \sqrt{\gamma(1-t)} \lvert 1000 \rangle_{\tilde{I}\tilde{E}\tilde{A}\tilde{F}}- \beta \sqrt{\gamma t} \lvert 1011 \rangle_{\tilde{I}\tilde{E}\tilde{A}\tilde{F}} \nonumber \\ & +\beta \sqrt{(1-\gamma)(1-t)} \lvert 0100 \rangle_{\tilde{I}\tilde{E}\tilde{A}\tilde{F}}+\beta\sqrt{t(1-\gamma)} \lvert 0111 \rangle_{\tilde{I}\tilde{E}\tilde{A}\tilde{F}}, \nonumber \\
     &(\mathrm{H}_{\tilde{I}}\otimes \mathrm{H}_{\tilde{A}})\lvert \Phi^{{\rm{(out)}}} \rangle \nonumber \\
    &= \alpha\sqrt{1-t} \lvert +0+0\rangle_{\tilde{I}\tilde{E}\tilde{A}\tilde{F}}+\alpha\sqrt{t}\lvert +0-1\rangle_{\tilde{I}\tilde{E}\tilde{A}\tilde{F}} \nonumber \\ &+\beta \sqrt{\gamma(1-t)} \lvert -0+0 \rangle_{\tilde{I}\tilde{E}\tilde{A}\tilde{F}}- \beta \sqrt{\gamma t} \lvert -0-1 \rangle_{\tilde{I}\tilde{E}\tilde{A}\tilde{F}} \nonumber \\ & +\beta \sqrt{(1-\gamma)(1-t)} \lvert +1+0 \rangle_{\tilde{I}\tilde{E}\tilde{A}\tilde{F}} \nonumber \\
    &+\beta\sqrt{t(1-\gamma)} \lvert +1-1 \rangle_{\tilde{I}\tilde{E}\tilde{A}\tilde{F}},
    \label{eq:ONENLAOUT}
\end{align}
where in the last step a change of basis was applied to modes $\tilde{I}$ and $\tilde{A}$ by performing two Hadamard operations. It can be seen from the previous expression that $\lvert \Phi^{{\rm{(out)}}} \rangle$ is a four mode entangled state. Performing two X-basis projective measurements on $\tilde{I}$ and $\tilde{A}$, disentangles them and directly entangles $\tilde{E}$ and $\tilde{F}$. The state's form is dependent on the outcomes of the projective measurements. 

When the measurement outcomes are similar, we denote this as operating point one ($\mathrm{OP}1$), the output state is  
\begin{align}
    \lvert \Phi^{({\rm{out}})}_{+}\rangle &= \frac{1}{\sqrt{N_{+}}} \Big[(\alpha \sqrt{1-t} \lvert 0 \rangle_{\tilde{F}}-\beta \sqrt{\gamma t}\lvert 1 \rangle_{\tilde{F}})\otimes \lvert 0 \rangle_{\tilde{E}} \nonumber \\ &+ \beta \sqrt{(1-\gamma)(1-t)} \lvert 0 \rangle_{\tilde{F}} \otimes \lvert 1 \rangle_{\tilde{E}}\Big],
    \label{eq:AmpSim}
\end{align}
where $N_{+} = \alpha^{2}(1-t)+\beta^{2} [\gamma t +(1-\gamma)(1-t)]$ is a normalisation constant.

The transformation of the 4-node cluster state into a 2-node one re-scaled the weights of the vacuum and single photon components. In order to quantify the amount of re-scaling, a gain expression is defined. By calculating the ratio between the single photon component in the final mode, $\Tilde{F}$, given a successful BSM, and the single photon component entering the amplifier, $\Tilde{I}$, we arrive at the desired gain expression \cite{osorio2012heralded, bruno2016heralded}
  \begin{align}
       G_{\rm{OP1}}=\frac{t}{\alpha^{2}(1-t)+\beta^{2}[\gamma t+(1-\gamma)(1-t)]}.
       \label{eq:GainSame}
  \end{align}

It can be seen from the previous expression that $G_{\rm{OP1}}>1$ when $t>1/2$. When $t=1/2$, $G_{\rm{OP1}}=1$ and the device operates as a teleporter. Furthermore, ideal gain, $G_{\rm{OP1}}\approx t/(1-t)$ \cite{ralph2009nondeterministic}, is achieved given a successful BSM and high input losses, i.e., $\gamma = 1$.
  
On the other hand, in the event of different measurement outcomes, denoted as operating point 2 ($\mathrm{OP}2$), the output state becomes 
\begin{align}
    \lvert \Phi^{({\rm{out}})}_{-}\rangle &= \frac{1}{\sqrt{N_{-}}} \Big[(\alpha \sqrt{t} \lvert 1 \rangle_{\tilde{F}}-\beta \sqrt{ \gamma(1- t)}\lvert 0 \rangle_{\tilde{F}})\otimes \lvert 0 \rangle_{\tilde{E}} \nonumber \\ &+ \beta \sqrt{t(1-\gamma)} \lvert 1 \rangle_{\tilde{F}} \otimes \lvert 1 \rangle_{\tilde{E}}\Big],
    \label{eq:AmpDiff}
\end{align}
where $N_{-} = \alpha^{2} t + \beta^{2}[\gamma(1-t)+t(1-\gamma)]$. 

As can be seen the weightings of the single photon and vacuum components are switched. This can be redeemed by performing an error correction step. First observe how the state transforms after applying a Pauli-X correction to the final mode
\begin{align}
    \lvert \Phi^{{\rm{(out)}}}_{-} \rangle &= \frac{1}{\sqrt{N_{-}}} \Big[(\alpha \sqrt{t} \lvert 0 \rangle_{\tilde{F}}-\beta \sqrt{ \gamma(1- t)}\lvert 1 \rangle_{\tilde{F}})\otimes \lvert 0 \rangle_{\tilde{E}} \nonumber \\ &+ \beta \sqrt{t(1-\gamma)} \lvert 0 \rangle_{\tilde{F}} \otimes \lvert 1 \rangle_{\tilde{E}}\Big].
\end{align}
\begin{figure}
     \centering
   \subfloat[]{\includegraphics[scale=0.18]{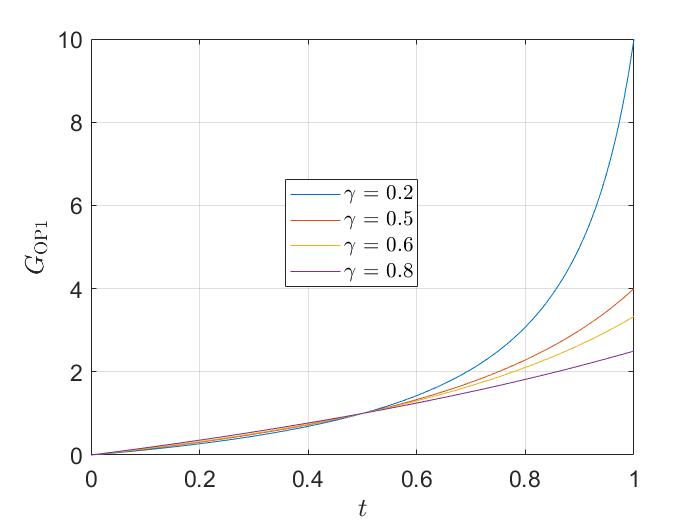}}
   \subfloat[]{\includegraphics[scale=0.18]{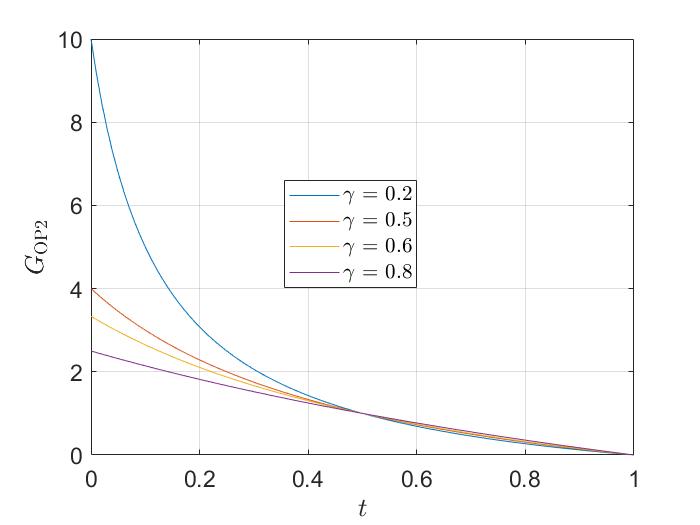}}
    \caption{A plot of the amplifier's gain against its tunable parameter $t$ for OP1 and OP2. In OP1, increasing the value of $t$, higher gain values can be achieved in order to restore an attenuated qubit initially prepared in a balanced superposition state, i.e., $\alpha = \beta = 1/\sqrt{2}$. On the other hand, by lowering the value of $t$, OP2 can achieve higher gain values. Effectively lowering the value $t$ in OP2 means higher reflectivity, while increasing $t$ in OP1 means high transmissivity and lower reflectivity. The gain is state dependent as depicted, it depends on how much attenuation is introduced by the transmitting channel and the qubit's initial weightings. We have considered different channel transmissivities and their corresponding gain curves. As can be seen gain is plotted for $\gamma = 0.2, 0.5, 0.6$, and $0.8$.  }
    \label{fig:Gaincurve}
\end{figure}
The corresponding gain expression in this case is 
\begin{align}
    G_{\rm{OP2}} = \frac{1-t}{\alpha^{2}t+\beta^{2}[\gamma(1-t)+t(1-\gamma)]}.
\end{align}

Contrary to the previous case, here a $G_{\rm{OP2}}>1$ operation is achieved when $t<1/2$, whereas when $t=1/2$ the device operates as a teleporter as before, and the ideal gain in this case would be $G_{\rm{OP2}} \approx (1-t)/t$.

The key feature of our device is that the BSM required for a successful amplification event was achieved by a CZ operation followed by two qubit X-basis measurements. This approach, as demonstrated in details, doesn't necessarily require photon counters in order to achieve amplification event. This is due to the fact that the device's final measurement is performed individually on each qubit, unlike a QS-NLA which detects a photon number interference event between two modes and thus requires photon counters to post-select the desired amplified output. Furthermore, the extra redundancy in our device, opened the way for an alternative operating point when a BSM fails, which can not be replicated by a QS-NLA.    

We end this section by analyzing the gain expressions of our device for both OP1 and OP2. In  Fig. (\ref{fig:Gaincurve}) we have plotted the amplifier's gain against its tunable parameter $t$. We firstly observe that both gain expressions are state dependent, i.e., they rely on $\alpha$, $\beta$, and channel loss $\gamma$. Thus, without loss of generality, we have assumed for both OPs in Figs. (3a) and (3b) respectively, that $\alpha=\beta=1/ \sqrt{2}$. In OP1, by increasing the transmissivity $t$, higher gain values can be achieved in order to restore an attenuated qubit. On the other hand, OP2 has higher tunable gain when the beamsplitter's reflectivity increases, in other words, when its transmissivity decreases.

We now turn our attention towards assessing the two models, QS-NLA and one-way NLA, against eachother by considering detection imperfections. We show in the next section the regime of operation where our device has a higher success probability and hence outperform a QS-NLA. 
\subsection{Effect of detector's noise and inefficiency}
In the previous sections it was pointed out that the state of the art QS-NLAs rely on (PNRDs) for successful amplification. As of today, microwave single photon counters remain elusive. Nonetheless, microwave photo-detectors have enjoyed great success, specifically, the \textit{quantum non-demolition } (QND) type, owing to the versatility of \textit{circuit quantum electrodynamics} (cQED) platforms. Practically, detection errors such as, non-ideal efficiency and dark count probability, would affect the performance of any QIP protocol. Hence, we focus in this section on studying these effects. Firstly, we begin with a brief description of the working principle of QND detectors, followed by a mathematical definition of the detection operators and then finally an evaluation of the success probabilities corresponding to the two possible device operating points. The last part of this section is dedicated to the performance analysis of the one-way NLA, specifically, we study  the relation between the device's success probability and gain for each operating point. Furthermore, we compare our device to a practical QS-NLA and show its capacity to achieve a higher success probability.   

Suppose that our device performs the X-basis projective measurements required for a successful amplification via QND detectors. Physically a QND detector is a \textit{qubit-cavity} system engineered such that the interaction between the qubit and a single photon populating the cavity is described as, $H_{\mathrm{int}} =i \Omega ( a \sigma^{+}+a^{\dagger} \sigma^{-}) $, where $\Omega$ is the qubit's Rabi frequency, and $\sigma^{+}(\sigma^{-})$, $a^{\dagger}(a)$ are the qubit and photon raising (lowering) operators respectively \cite{haroche2006exploring}.

When the cavity captures an arbitrary single photon wave-packet $\alpha \lvert 0 \rangle+ \beta \lvert 1 \rangle$, a series of controlled qubit pulses evolve the compound system into an entangled state that can be written as $\alpha \lvert 0g\rangle + \beta \lvert 1e\rangle$. As described earlier, applying a Hadamard transformation to the photonic degree of freedom transforms the state into, $\alpha \lvert +g \rangle+ \beta \lvert -e \rangle$. When the single photon state is $\lvert + \rangle$, the qubit assumes its ground state, whereas when the photon state is $\lvert - \rangle $ the qubit is in the excited state. Thus, the presence of a photon in any of the 2-dimensional diagonal basis states, $\{\lvert+\rangle, \lvert-\rangle\}$, is detectable by measuring the qubit state. 

The aforementioned QND detector provides information on the presence or absence of a specific single photon state. This process is denoted as \textit{on-off} detection  \cite{PhysRevA.86.042328, PhysRevA.93.042328, petrovnin2023microwave}. The measurement operators describing an ideal on-off detector are defined as 
\begin{align}
    \mathrm{M}_{\mathrm{off}} &= \lvert 0 \rangle \langle 0 \lvert \nonumber \\ 
    \mathrm{M}_{\mathrm{on}} &= \mathrm{I}- \mathrm{M}_{\mathrm{off}}= \underset{n=1}{\overset{\infty}{\sum}} \lvert n \rangle \langle n \lvert.
\end{align}

For the case at hand, the QND detector "on" operator, $\mathrm{M}_{\mathrm{on}}$, is truncated to the single photon level. Detector imperfections manifest as \textit{dark count probability}, which is incorporated to the off operator, $M_{\rm{off}}$, and inefficiency, which is modeled by a beamsplitter of transmissivity $\eta$ followed by a perfect detector. More precisely, the mode to be detected impinges first on an attenuating beamsplitter then followed by an on-off detector, $a_{\rm{out}} = \sqrt{\eta}a_{\rm{in}}+\sqrt{1-\eta}a_{v}$, where $a_{\rm{in}}$ is the input to be detected, $a_{\rm{out}}$ is the beamsplitter's output, and $a_{v}$ is a vacuum mode entering from the beamsplitter's unused port. In addition to that, we recast the photon projector operators in the Hadamard basis. Thus the on-off operators become
\begin{align}
    \mathrm{M}_{\mathrm{off}} &= (1-\mu) \lvert + \rangle \langle + \lvert \nonumber \\ 
    \mathrm{M}_{\mathrm{on}} &=  \mu \lvert + \rangle \langle + \lvert \hspace*{0.5em} + \hspace*{0.5em} \lvert - \rangle \langle - \lvert, 
\end{align}
where $\mu$ is the dark count probability.

Let us now focus on the output state of our device, Eq. (\ref{eq:ONENLAOUT}). In our scheme, two QND on-off detectors are utilised to perform a projective measurement on the two middle cluster nodes. A successful amplification event is decided based on measuring the following observables
\begin{align}
    \mathrm{M}_{0} &= \mathrm{M}_{\mathrm{off}} \otimes \mathrm{M}_{\mathrm{off}} \nonumber \\ 
    \mathrm{M}_{1} &= \mathrm{M}_{\mathrm{off}} \otimes \mathrm{M}_{\mathrm{on}} \nonumber \\ 
    \mathrm{M}_{2} & = \mathrm{M}_{\mathrm{on}} \otimes \mathrm{M}_\mathrm{off} \nonumber \\ 
    \mathrm{M}_{3} &= \mathrm{M}_{\mathrm{on}} \otimes \mathrm{M}_{\mathrm{on}},
\end{align}
where an "off" event corresponds to measuring a photon in the $\lvert +\rangle$ state, whereas an "on" event corresponds to measuring a $\lvert - \rangle$ state. After recasting Eq. (\ref{eq:ONENLAOUT}) as follows
\begin{align}
    &\lvert \Phi^{{\rm{(out)}}} \rangle \nonumber \\
    &= \alpha\sqrt{1-t} \lvert 00\rangle_{\tilde{E}\tilde{F}} \otimes \lvert \Psi_{0} \rangle+\alpha\sqrt{t}\lvert 01\rangle_{\tilde{E}\tilde{F}} \otimes \lvert \Psi_{1}\rangle \nonumber \\ &+\beta \sqrt{\gamma(1-t)} \lvert 00 \rangle_{\tilde{E}\tilde{F}} \otimes \lvert \Psi_{2} \rangle- \beta \sqrt{\gamma t} \lvert 01 \rangle_{\tilde{E}\tilde{F}}\otimes \lvert \Psi_{3} \rangle \nonumber \\ & +\beta \sqrt{(1-\gamma)(1-t)} \lvert 10 \rangle_{\tilde{E}\tilde{F}} \otimes \lvert \Psi_{0} \rangle\nonumber \\
    &+\beta\sqrt{t(1-\gamma)} \lvert 11 \rangle_{\tilde{E}\tilde{F}}\otimes \lvert \Psi_{1} \rangle,
\end{align}
where $\lvert \Psi_{0} \rangle = \lvert ++\rangle$, $\lvert \Psi_{1} \rangle = \lvert +-\rangle$, $\lvert \Psi_{2} \rangle = \lvert -+\rangle$, and $\lvert \Psi_{3} \rangle = \lvert --\rangle$.

The probability of successfully obtaining a specific combination of the aforementioned on-off observables becomes
\begin{align}
P(l) &=   \langle \Phi^{{\rm{(out)}}} \lvert \mathrm{I} \otimes \mathrm{M}_{l} \otimes \mathrm{I} \lvert \Phi^{{\rm{(out)}}} \rangle,
\end{align}
where $l=0,1,2,3$.

Consequently the success probabilities can be calculated as
\begin{align}
    &\langle \Phi^{{\rm{(out)}}} \lvert \mathrm{I} \otimes \mathrm{M}_{0} \otimes \mathrm{I} \lvert \Phi^{{\rm{(out)}}} \rangle \nonumber \\ 
    &= (1-\mu)^{2} [(1-t)[\alpha^{2}+\beta^{2}(1-\gamma)]] \nonumber  \\
    &\langle \Phi^{{\rm{(out)}}} \lvert \mathrm{I} \otimes \mathrm{M}_{1} \otimes \mathrm{I} \lvert \Phi^{{\rm{(out)}}} \rangle \nonumber \\ &= (1-\mu)[\mu(1-\eta)+\eta ][\alpha^{2}t+\beta^{2}t(1-\gamma)] \nonumber \\ 
    &\langle \Phi^{{\rm{(out)}}} \lvert \mathrm{I} \otimes \mathrm{M}_{2} \otimes \mathrm{I} \lvert \Phi^{{\rm{(out)}}} \rangle \nonumber \\
    &= (1-\mu)[\mu(1-\eta)+\eta ] [\beta^{2}\gamma (1-t)] \nonumber \\ 
    &\langle \Phi^{{\rm{(out)}}} \lvert \mathrm{I} \otimes \mathrm{M}_{3} \otimes \mathrm{I} \lvert \Phi^{{\rm{(out)}}} \rangle \nonumber \\
    &= \beta^{2}\gamma t [\mu(1-\eta)+\eta ]^{2}.
    \label{eq:ProbEvent}
\end{align}
where detection inefficiency was manifested as the following transformation, $\lvert - \rangle = \sqrt{\eta} \lvert - \rangle + \sqrt{1-\eta} \lvert + \rangle $.

According to the previous equation the device has two success probabilities corresponding to its two operating points. When $\mathrm{OP1}$ is true, the probability of success $P^{\mathrm{succ}}_{\rm{OP1}}$ corresponds to the sum of on-on and off-off events. On the other hand, when $\mathrm{OP}2$ is true, the success probability $P^{\mathrm{succ}}_{\rm{OP2}}$ corresponds to the on-off and off-on events after error correction.

In practice the efficiency and dark count probability of a QND detector are $\eta \approx 0.75-0.85$, and $\mu \approx 0.015$ respectively \cite{johnson2010quantum, PhysRevX.8.021003, kono2018quantum}. However, due to TM mismatch errors, the efficiency may drop to $50 \%$ \cite{PhysRevX.6.031036}. We consider both cases in the next section and show the effect of reducing the detector's efficiency on the success probability of our device. We also compare the performance of our device to a QS-NLA similar to the one depicted in Fig. (\ref{fig:NLA}) utilizing PNRDs, and show that our device has an enhanced success probability besides being fault-tolerant.  
\subsection{Performance analysis}
\begin{figure*}
    \centering
   \subfloat[]{\includegraphics[scale=0.24]{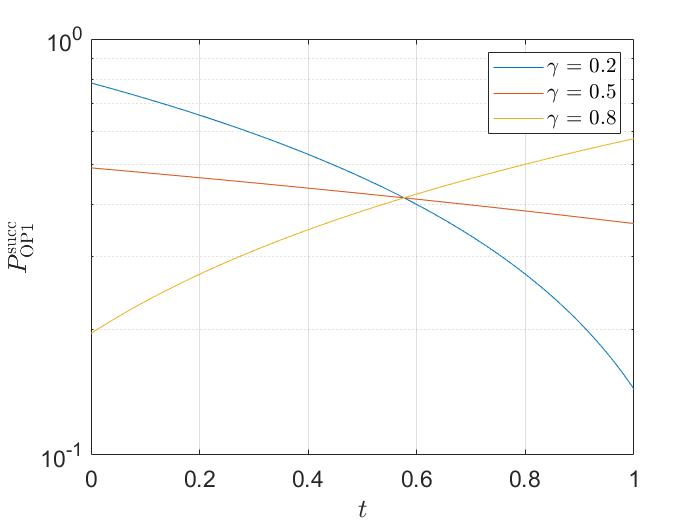}}
   \subfloat[]{\includegraphics[scale=0.24]{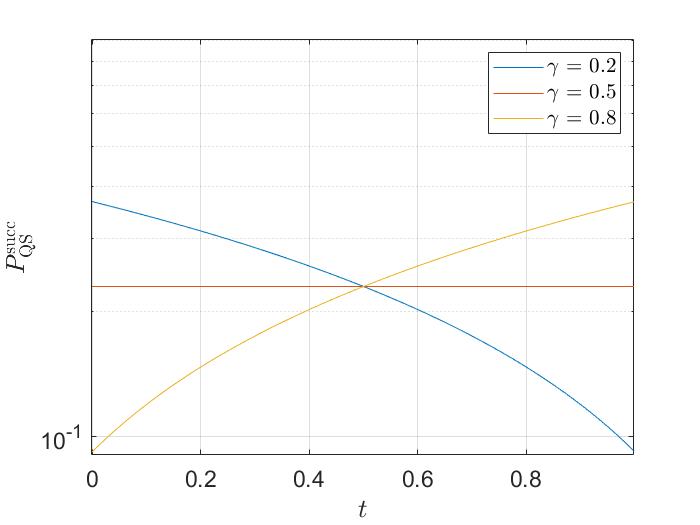}}
   \subfloat[]{\includegraphics[scale=0.24]{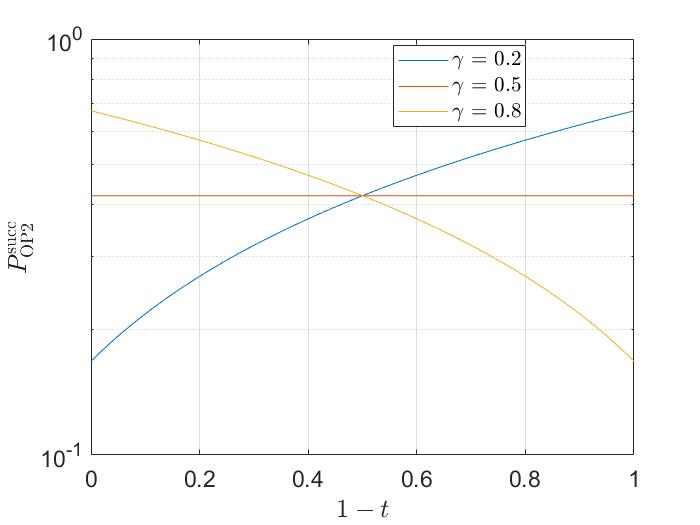}}
   \\
   \subfloat[]{\includegraphics[scale=0.24]{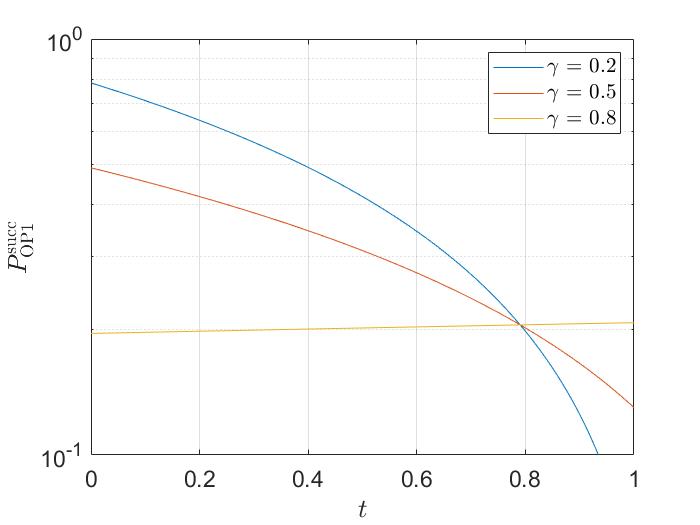}}
   \subfloat[]{\includegraphics[scale=0.24]{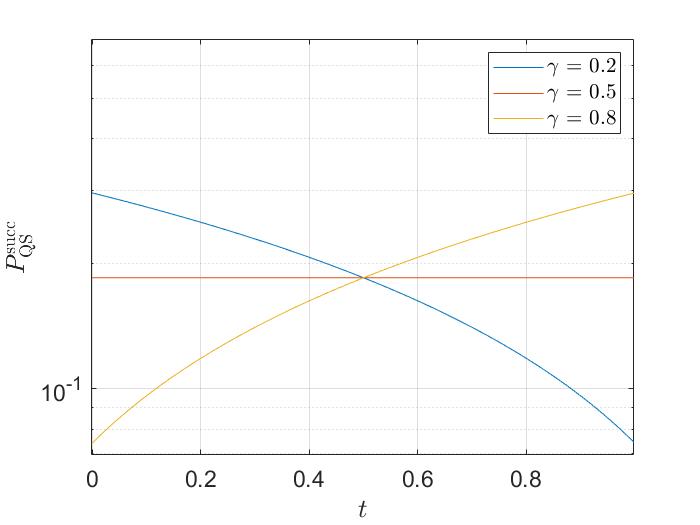}}
   \subfloat[]{\includegraphics[scale=0.24]{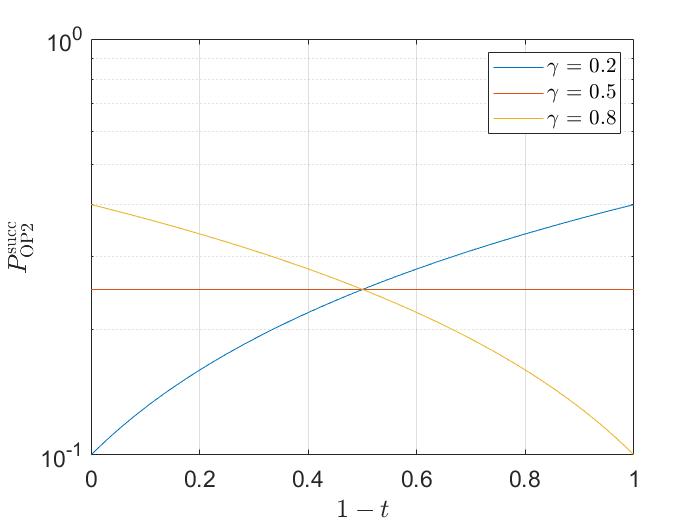}}
    \caption{Success probability curves of one-way NLAs and QS-NLA versus their tuning parameter $t$ respectively. The channel's input was assumed to be a single photon, i.e., $\alpha=0$, $\beta=1$. In the above panels detector's efficiency is $\eta \approx 0.85$, whereas in the below ones is $\eta \approx 0.5$. A dark count probability, $\mu \approx 0.015$, was assumed for both one-way NLAs, whereas $\mu \approx 0.5$ for QS-NLA. An efficieny of $\eta \approx 0.5$ was considered for all amplifiers in the below panels, while dark count probability is the same as before respectively. Figs. (4a) and (4c) show that both one-way NLAs outperform a QS-NLA for different channel transmissivities respectively. In Fig. (4a) OP1's success probability is controlled by the transmissivity of the auxiliary photon, whereas OP2's success probability is controlled by its reflectivity. OP1 and OP2 outperform QS-NLA for all the considered channel transmissivites except for the case when $\gamma = 0.8$ where QS-NLA outperforms OP1.}
    \label{fig:NLAPsuccess}
\end{figure*}
 Fig. (\ref{fig:NLAPsuccess}) is a plot of the success probability of a one-way NLA against its tunable parameter for different detector efficiencies, channel transmissivites and operating points. The plot also includes a QS-NLA in order to compare the two models against eachother. Without loss of generality, we have assumed a single photon as the channel's input, i.e., $\alpha = 0$, $\beta=1$. In the upper panel Figs. (4a), (4b), and (4c) consider one-way NLA OP1, QS-NLA, and one-way NLA OP2 respectively. Both one-way NLAs utilize QND detectors whereas QS-NLA utilize PNRDs. The detection efficiency was assumed $\eta=0.85$ for all detectors. Dark count probability for QND detectors was assumed $\mu=0.015$, while that for PNRDs is $\mu=0.5$. This is due to the fact that ideal microwave photon counters  are currently unavailable \cite{casariego2022propagating}. As can be seen, both one-way NLAs outperform QS-NLA in terms of the probability of success. This is due to the fact that a BSM in one-way NLA heralds a successful amplification event by performing two separate photodetections. Unlike QS-NLA where a BSM rely on detecting a two mode photon number interference pattern with PNRDs. For one-way OP1 and QS-NLA, when the channel's transmissivity is $\gamma<1/2$, the success probability decreases by increasing $t$. This can be justified by observing that  when the input's weightings are already in favor of the vaccum component, a large value of $t$ means that the auxilary photon is mostly transmitted and a BSM fails since with high probability no photons enter the bell state analyzer. On the other hand when $\gamma>1/2$ the success probability increases by increasing $t$, since in this case the input weightings are already in favor of the single photon component and by increasing $t$ less fraction of the auxiliary photon interferes with the input and a BSM succeeds since with high probability only one photon enters the bell state analyzer. For one-way NLA OP2, the amplifier's gain is controlled by varying the auxiliary beamsplitter reflectivity $1-t$ (see Fig. (\ref{fig:Gaincurve})). As can be seen from Fig. (4c) when $\gamma<1/2$ the success probability increases by increasing the reflectivity, since the input weightings are already in favor of the vacuum mode and large reflectivity means a larger portion of the auxiliary mode enters the amplifier and hence enhances the success of a BSM. On the other hand when $\gamma>1/2$, i.e., weightings are in favor of the single photon component,  the success probability decreases by increasing the reflectivity, since this situation leads to a larger portion of the auxiliary photon entering the amplifier and hence a BSM fails. 
 In the lower panel the same amplifiers were considered, however the detection efficiency was assumed $\eta=0.5$ for all detectors. The dark count probability was assumed as before respectively. Similarly it can be seen that both one-way OP1 and OP2 outperform QS-NLA for all channel transmissivities, except when $\gamma = 0.8$, it can be seen that QS-NLA can only beat OP1 while being outperformed by OP2.
\section{APPLICATIONS OF ONE-WAY NLA}
In this section we are mainly concerned with microwave quantum sensing applications that thrive on low power signals. Quantum illumination \cite{lloyd2008enhanced, tan2008quantum}, and entanglement-enhanced radar sensing \cite{lanzagorta2012quantum} are the most studied protocols fitting this category. 

The aforementioned protocols utilize an entanglement resource in the form of a TMSV in order to interrogate a target. A probe denoted as \textit{signal} is sent towards a target, whereas its twin (denoted as \textit{idler}) is retained in a lossy memory element for a correlation measurement with the target's return. Analogously, we can regard the dissipative dynamics of the stored mode as a propagation through an ADC. Thus our objective here is to evaluate the cost at which NLAs are capable of protecting an idler mode against storage losses during an entanglement-based quantum sensing protocol.

\subsection{Low power microwave quantum sensing }
Suppose now that a TMSV is generated at a transmitter
\begin{align}
    \lvert \Psi \rangle_{SI} = \underset{n=0}{\overset{\infty}{\sum }}\sqrt{\frac{N^{n}_{S}}{(N_{S}+1)^{n+1}}} \lvert n \rangle_{S}\lvert n\rangle_{I},
\end{align}
where $N_{S}$ is the mean photon number in each of the signal and idler modes, i.e., $\langle a^{\dagger}_{S}a_{S} \rangle=\langle a^{\dagger}_{I}a_{I}\rangle =N_{S}$. It is often more convenient to recast the previous expression by  defining the following parameter $\lambda^{2} = \frac{N_{S}}{1+N_{S}}$, hence
\begin{align}
     \lvert \Psi \rangle_{SI} = \sqrt{1-\lambda^{2}} \underset{n=0}{\overset{\infty}{\sum }}\lambda^{n} \lvert n \rangle_{S}\lvert n\rangle_{I}.
\end{align}
Assume now that the idler mode is undergoing an ADC representing a lossy memory element with transmissivity $\gamma$. Similarly as before, this can be modeled by a passive beamsplitter with the same transmissivity, where its unused port injects vacuum. In the Heisenberg picture, the corresponding beamsplitter unitary operator can be defined as, $U_{\theta}=e^{\theta(a^{\dagger}b+b^{\dagger}a)}$, where the beamsplitter's transmissivity equals $\gamma= \cos^{2}{\theta}$, and $a$, $b$ are its modes. Thus the input transforms as 
\begin{align}
    \lvert \Psi \rangle_{SI} &= \sqrt{1-\lambda^{2}}\underset{n=0}{\overset{\infty}{\sum}} \lambda^{n} \frac{{a^{\dagger}_{S}}^{n}}{\sqrt{n!}} \frac{{a^{\dagger}_{I}}^{n}}{\sqrt{n!}}  \lvert 0 \rangle^{\otimes 3}_{SIE} \nonumber \\ 
   U_{\theta}\lvert \Psi \rangle_{SI} &= \sqrt{1-\lambda^{2}}\underset{n=0}{\overset{\infty}{\sum}} \lambda^{n} \frac{{a^{\dagger}_{S}}^{n}}{\sqrt{n!}} \frac{[{U_{\theta}a^{\dagger}_{I}U_{\theta}^{\dagger}]}^{n}}{\sqrt{n!}}  \lvert 0 \rangle^{\otimes 3}_{SIE} \nonumber \\ 
    &=\sqrt{1-\lambda^{2}}\underset{n=0}{\overset{\infty}{\sum}} \lambda^{n} \frac{{a^{\dagger}_{S}}^{n}}{\sqrt{n!}} \frac{[{\sqrt{\gamma} a^{\dagger}_{I}+\sqrt{1-\gamma}a^{\dagger}_{E} ]}^{n}}{\sqrt{n!}} \lvert 0 \rangle^{\otimes 3}_{SIE}. \nonumber
\end{align}
By utilizing the Binomial expansion theorem we get
\begin{align}
    \lvert \Psi \rangle_{SIE} &= \sqrt{1-\lambda^{2}}\underset{n=0}{\overset{\infty}{\sum}} \lambda^{n}  \underset{k=0}{\overset{n}{\sum}}{n \choose k}\frac{{a^{\dagger}_{S}}^{n}}{\sqrt{n!}} \nonumber \\ 
    &\frac{[\gamma^{\frac{n-k}{2}}{a^{\dagger}_{I}}^{n-k}][(1-\gamma)^{\frac{k}{2}}{a^{\dagger}_{E}}^{k}]}{\sqrt{n!}} \lvert 0 \rangle^{\otimes 3}_{SIE}, \nonumber
\end{align}
where $\lvert 0 \rangle^{\otimes 3}_{SIE} = \lvert 0 \rangle_{S} \otimes \lvert 0 \rangle_{I} \otimes \lvert 0 \rangle_{E}$, and $E$ is an environment mode.

Then finally by utilizing the properties of the factorial function we arrive at
\begin{align}
    \lvert \Psi \rangle_{SIE} &= \sqrt{1-\lambda^{2}}\underset{n=0}{\overset{\infty}{\sum}} \lambda^{n}  \underset{k=0}{\overset{n}{\sum}} \sqrt{{n \choose k}} \gamma^{\frac{n-k}{2}}(1-\gamma)^{\frac{k}{2}} \nonumber \\ & \lvert n \rangle_{S} \lvert n-k\rangle_{I} \lvert k \rangle_{E}. 
\end{align}
For low powered quantum sensing applications, the number of photons in each of the signal and idler modes is approximately $N_{S} \ll 1$. Thus we can truncate the previous expansion to first order
\begin{align}
    \lvert \Psi \rangle_{SIE} & \approx \sqrt{1-\lambda^{2}} \big[\lvert  000\rangle_{SIE} + \lambda \sqrt{\gamma} \lvert 110 \rangle_{SIE} \nonumber \\ &+ \lambda \sqrt{1-\gamma} \lvert 101\rangle_{SIE} \big] \nonumber \\ 
    &\approx \alpha \lvert 000 \rangle_{SIE} + \beta \sqrt{\gamma} \lvert 110 \rangle_{SIE} + \beta \sqrt{1-\gamma} \lvert 101\rangle_{SIE},
    \label{eq:WeakTMSV}
\end{align}
where $\alpha = \sqrt{1-\lambda^{2}}$, $\beta = \lambda \sqrt{1-\lambda^{2}}$.

In order to complete the amplification protocol, the receiver prepares an auxiliary mode defined as 
\begin{align}
    \lvert \Psi \rangle_{AF} = \sqrt{1-t} \lvert 00 \rangle_{AF}+\sqrt{t} \lvert 11 \rangle_{AF}.
\end{align}
Thus the whole state of the amplifier becomes
\begin{align}
    \lvert \Psi \rangle _{SIEAF} &=\lvert \Psi \rangle_{SIE} \otimes \lvert \Psi \rangle_{AF}.
\end{align}
After that modes $I$ and $A$ are redirected towards a CZ gate, such that, mode $I$ is the control, whereas mode $A$ is the target. We note that exchanging the roles of the control and target modes won't affect the protocol's outcome due to the symmetry of the CZ gate 
\begin{align}
   \lvert \Psi \rangle_{ \tilde{S}\tilde{I}\tilde{E}\tilde{A}\tilde{F}} &= \alpha \sqrt{1-t} \lvert0+0+0 \rangle_{ \tilde{S}\tilde{I}\tilde{E}\tilde{A}\tilde{F}} \nonumber \\ &+ \alpha\sqrt{t}\lvert 0+0-1\rangle_{ \tilde{S}\tilde{I}\tilde{E}\tilde{A}\tilde{F}} \nonumber \\ &+\beta \sqrt{\gamma(1-t)} \lvert 1-0+0 \rangle_{ \tilde{S}\tilde{I}\tilde{E}\tilde{A}\tilde{F}} \nonumber \\ &-\beta \sqrt{\gamma t} \lvert 1-0-1 \rangle_{ \tilde{S}\tilde{I}\tilde{E}\tilde{A}\tilde{F}} \nonumber \\ &+ \beta \sqrt{(1-\gamma)(1-t)} \lvert 1+1+0 \rangle_{ \tilde{S}\tilde{I}\tilde{E}\tilde{A}\tilde{F}} \nonumber \\&+ \beta \sqrt{t(1-\gamma)} \lvert 1+1-1\rangle_{ \tilde{S}\tilde{I}\tilde{E}\tilde{A}\tilde{F}},
\end{align}
where $\tilde{I}$ and $\tilde{A}$ are written in the diagonal basis. 

Then two X-basis measurements are performed on modes $\tilde{I}$ and $\tilde{A}$, yielding the following output state when the measurement outcomes are the same 
\begin{align}
    \lvert \Psi \rangle_{ \tilde{S}\tilde{E}\tilde{F}} &= \frac{1}{\sqrt{N_{+}}}\Big[(\alpha \sqrt{(1-t)} \lvert 00 \rangle_{\tilde{S}\tilde{F}}- \beta \sqrt{\gamma t} \lvert 11\rangle_{\tilde{S}\tilde{F}})\otimes \lvert 0\rangle_{\tilde{E}} \nonumber \\ &+ \beta \sqrt{(1-\gamma)(1-t)} \lvert 10\rangle_{\tilde{S}\tilde{F}}\otimes \lvert1\rangle_{\tilde{E}}\Big].
    \label{eq:SensingSame}
\end{align}

Conversely, when the outcomes are different we get
\begin{align}
    \lvert \Psi \rangle_{ \tilde{S}\tilde{E}\tilde{F}} &= \frac{1}{\sqrt{N_{-}}}\Big[(\alpha \sqrt{t} \lvert 01 \rangle_{\tilde{S}\tilde{F}}- \beta \sqrt{\gamma (1-t)} \lvert 10\rangle_{\tilde{S}\tilde{F}})\otimes \lvert 0\rangle_{\tilde{E}} \nonumber \\ &+ \beta \sqrt{(1-\gamma)t} \lvert 11\rangle_{\tilde{S}\tilde{F}}\otimes \lvert1\rangle_{\tilde{E}}\Big],
    \label{eq:SensingDiff}
\end{align}
where $N_{+}$ and $N_{-}$ are normalization constants defined as before. 

Then we can operate on the erroneous state in Eq. (\ref{eq:SensingDiff}) by applying a Pauli X correction on mode $\tilde{F}$ as described earlier in details (see section \Romannum{3}B). Thus, the amplifier's output state for each case respectively becomes 
\begin{align}
    \rho^{(\rm{out})}_{+} &= \frac{1}{N_{+}} \Big [ \lvert \Psi^{\rm{(out)}}_{+}\rangle \langle \Psi^{\rm{(out)}}_{+} \lvert + \beta^{2}(1-\gamma)(1-t) \lvert 10 \rangle \langle 01\lvert \Big],
\end{align}
where $\lvert \Psi^{\rm{(out)}}_{+} \rangle = \frac{1}{\sqrt{N_{+}}}\big[\alpha \sqrt{1-t} \lvert 00 \rangle_{\tilde{S}\tilde{F}} - \beta \sqrt{\gamma t} \lvert 11 \rangle_{\tilde{S}\tilde{F}}\big]$
\begin{align}
    \rho^{(\rm{out})}_{-} &= \frac{1}{N_{-}} \Big [ \lvert \Psi^{\rm{(out)}}_{-}\rangle \langle \Psi^{\rm{(out)}}_{-} \lvert + \beta^{2}(1-\gamma) t \lvert 10 \rangle \langle 01\lvert \Big],
\end{align}
where $\lvert \Psi^{\rm{(out)}}_{-} \rangle = \frac{1}{\sqrt{N_{-}}}\big[\alpha \sqrt{t} \lvert 00 \rangle_{\tilde{S}\tilde{F}} - \beta \sqrt{\gamma (1-t)} \lvert 11 \rangle_{\tilde{S}\tilde{F}}\big]$.

Considering the previous equations more carefully, it can be seen that when the environment mode is in the vacuum state, the amplifier's output is an entangled state, whereas error occurs when the environment captures an input photon. As previously described in great details, by increasing (decreasing) $t$ for OP1(OP2), the probability amplitudes of the entangled output can be rescaled in favor of the single photon component, and thus restoring the lost photon.
%By tracing out the signal mode, $S$, it can be seen that the single photon component of the the final mode, $F$, is rescaled by the amplifiers tunable parameter $t$, %whereas, the vaccum is rescaled by its complement, and thus the desired amplification is achieved as before.  

We end this section by considering a practical entanglement-enhanced sensing scenario where a one-way NLA is utilized to restore a non-ideally stored idler mode $I$. 

Let's consider a microwave storage device of $ 50 \%$ efficiency. This corresponds to a lossy channel with transmissivity equal to $\gamma = 0.5$. The number of signal-idler probe pairs that can be generated in a typical microwave quantum sensing protocol is approximately $ M \approx 10^{5}-10^{6}$. Due to the probabilistic nature of NLAs, this number is rescaled by the device's success probability. Thus the number of generated pairs has to be increased in order to compensate for this reduction. Considering the worst case scenario first, it can be seen from Fig. (\ref{fig:NLAPsuccess}) that when $\mathrm{OP1}$ is true, and the detector's efficiency is $0.5$, the success probability corresponding to a complete restoration of the single photon component is $\approx 0.5$, whereas when $\mathrm{OP2}$ is true it becomes $0.24$. Since one-way NLA is fault tolerant and has uninterruptible operation, its success probability on average is $\approx 0.4$. On the other hand for the same case a QS-NLA has a success probability of $\approx 0.2$. Consequently the minimum number of required pairs would be $2.5 \times 10^{5}$ for one-way NLA, and $5 \times 10^{5}$ for QS-NLA which is double that of one-way NLA. For completeness we consider the case where the efficiency is $\eta \approx 0.85$, the average success probability of one-way NLA is approximately $\approx 0.5$ and that of QS-NLA is $\approx 0.3$ thus the minimum number of required pairs would be $\approx 2 \times 10^{5}$ for the former and $\approx 3.5 \times 10^{5}$ for the latter. 

To conclude, we have shown that a one-way NLA is capable of complete restoration of a lossy idler mode participating in a quantum sensing protocol. The corresponding cost was an increase in the number of utilized probe pairs, however compared to a QS-NLA our device requires a fewer number of pairs which would result in a faster quantum sensing protocol. 
\subsection{Entangling remote superconducting qubits}

We turn our attention towards the task of entangling two remote network nodes. As initially proposed, successful entanglement is established by performing a joint BSM in a mid-way station between the two communicating parties \cite{duan2001long}. Recently the intriguing idea of NLA has been deployed in order to implement this BSM \cite{ PhysRevA.102.052425, seshadreesan2020continuous,PhysRevA.107.042606}. 

When BS1 $\&$ BS2 in Fig. ~\ref{fig:NLA} are considered as two channels linking two remote nodes to a midway balanced beamsplitter, the resemblance between entanglement sharing and NLA protocols is clear. However, as pointed out earlier, heralded NLA requires 
\begin{figure}
    \centering
    \begin{tikzpicture}
        \draw [thick, fill= gray!20 ] (0,-0.3)--(0,0)--(1,0)--(1,-0.3);
        \draw[thick] (0,0)--(0,-0.3);
        \draw [thick, fill=white] plot [smooth] coordinates {(0,-0.3)(0.15, -0.15) (0.85,-0.15)(1,-0.3)};
        \shade[ball color = gray!40, opacity = 0.4] (0.5,-0.7) circle (0.25);
          \draw (0.5,-0.7) ellipse (0.25 and .07);
          \draw[ dotted] (0.5,-0.7) ellipse (0.1 and 0.25);
\draw [thick, fill= gray!20 ] (0,-1.1)--(0,-1.4)--(1,-1.4)--(1,-1.1);
        \draw [thick, fill=white] plot [smooth] coordinates {(0,-1.1)(0.15, -1.25) (0.85,-1.25)(1,-1.1)};

\draw [thick, fill= gray!20 ] (1.5,-0.3)--(1.5,0)--(2.5,0)--(2.5,-0.3);
        \draw[thick] (1.5,0)--(1.5,-0.3);
        \draw [thick, fill=white] plot [smooth] coordinates {(1.5,-0.3)(1.65, -0.15) (2.35,-0.15)(2.5,-0.3)};
        \shade[ball color = gray!40, opacity = 0.4] (2,-0.7) circle (0.25);
          \draw (2,-0.7) ellipse (0.25 and .07);
          \draw[ dotted] (2,-0.7) ellipse (0.1 and 0.25);
\draw [thick, fill= gray!20 ] (1.5,-1.1)--(1.5,-1.4)--(2.5,-1.4)--(2.5,-1.1);
        \draw [thick, fill=white] plot [smooth] coordinates {(1.5,-1.1)(1.65, -1.25) (2.35,-1.25)(2.5,-1.1)};

   \draw [thick, fill= gray!20 ] (3,-0.3)--(3,0)--(4,0)--(4,-0.3);
        \draw[thick] (3,0)--(3,-0.3);
        \draw [thick, fill=white] plot [smooth] coordinates {(3,-0.3)(3.15, -0.15) (3.85,-0.15)(4,-0.3)};
        \shade[ball color = gray!40, opacity = 0.4] (3.5,-0.7) circle (0.25);
        \draw[-stealth, thick] (3.5,-0.7)--(3.5,-0.45);
        \draw[-stealth, thick] (3.5,-0.7)--(3.5,-0.95);
        \draw[-stealth, thick] (0.5,-0.7)--(0.5,-0.45);
        \draw[-stealth, thick] (0.5,-0.7)--(0.5,-0.95);
        \draw[-stealth, thick] (2,-0.7)--(2,-0.45);
        \draw[-stealth, thick] (2,-0.7)--(2,-0.95);
        \draw[-stealth, thick] (5,-0.7)--(5,-0.45);
        \draw[-stealth, thick] (5,-0.7)--(5,-0.95);
          \draw (3.5,-0.7) ellipse (0.25 and .07);
          \draw[ dotted] (3.5,-0.7) ellipse (0.1 and 0.25);
\draw [thick, fill= gray!20 ] (3,-1.1)--(3,-1.4)--(4,-1.4)--(4,-1.1);
        \draw [thick, fill=white] plot [smooth] coordinates {(3,-1.1)(3.15, -1.25) (3.85,-1.25)(4,-1.1)};

     \draw [thick, fill= gray!20 ] (4.5,-0.3)--(4.5,0)--(5.5,0)--(5.5,-0.3);
        \draw[thick] (4.5,0)--(4.5,-0.3);
        \draw [thick, fill=white] plot [smooth] coordinates {(4.5,-0.3)(4.65, -0.15) (5.35,-0.15)(5.5,-0.3)};
        \shade[ball color = gray!40, opacity = 0.4] (5,-0.7) circle (0.25);
          \draw (5,-0.7) ellipse (0.25 and .07);
          \draw[ dotted] (5,-0.7) ellipse (0.1 and 0.25);
\draw [thick, fill= gray!20 ] (4.5,-1.1)--(4.5,-1.4)--(5.5,-1.4)--(5.5,-1.1);
\draw [thick, fill=white] plot [smooth] coordinates {(4.5,-1.1)(4.65, -1.25) (5.35,-1.25)(5.5,-1.1)};    
\draw[-stealth, thick, decorate,decoration={snake, amplitude=3.5pt, segment length = 6.5pt,pre length=2pt,post length=3pt}] (0.5,0)--(0.5,0.7);
\draw[-stealth, thick, decorate,decoration={snake, amplitude=3.5pt, segment length = 6.5pt,pre length=2pt,post length=3pt}] (2,0) --(2,0.7);
\draw[-stealth, thick, decorate,decoration={snake, amplitude=3.5pt, segment length = 6.5pt,pre length=2pt,post length=3pt}] (3.5,0) --(3.5,0.7);
\draw[-stealth, thick, decorate,decoration={snake, amplitude=3.5pt, segment length = 6.5pt,pre length=2pt,post length=3pt}] (5,0) --(5,0.7);
 \filldraw[fill=white] (0.5,0.95) circle (0.25); 
 \filldraw[fill=white] (2,0.95) circle (0.25);
 \filldraw[fill=white] (3.5,0.95) circle (0.25);
 \filldraw[fill=white] (5,0.95) circle (0.25);
 \draw[] (0.75, 0.95)--(1.75,0.95);
 \draw[] (1.25,1.2)--(1.25,1.2) node []{\scriptsize{$\rm CZ$}};
  \draw[] (2.7,1.2)--(2.7,1.2) node []{\scriptsize{$\rm CZ$}};
    \draw[] (4.3,1.2)--(4.3,1.2) node []{\scriptsize{$\rm CZ$}};
 \draw[thick,dotted] (2.25, 0.95)--(3.25,0.95);
 \draw[] (3.75, 0.95)--(4.75,0.95);
 \draw[](2,1.2)--(2,1.7);
 \draw[](3.5,1.2)--(3.5,1.7);
 \draw[](2,1.7)--(2.15,1.7)--(2.15,2)--(1.85,2)--(1.85,1.7)--(2,1.7);
 \draw[](3.5,1.7)--(3.65,1.7)--(3.65,2)--(3.35,2)--(3.35,1.7)--(3.5,1.7);
 \draw [] (2,1.85)--(2,1.85) node [] {\scriptsize $\mathrm{X}$};
  \draw [] (3.5,1.85)--(3.5,1.85) node [] {\scriptsize $\mathrm{X}$};
  \draw [] (0.5,0.95)--(0.5,0.95) node [] {\scriptsize $S$};
   \draw [] (2,0.95)--(2,0.95) node [] {\scriptsize $I$};
    \draw [] (3.5,0.95)--(3.5,0.95) node [] {\scriptsize $A$};
     \draw [] (5,0.95)--(5,0.95) node [] {\scriptsize $F$};
      \draw [] (2,0.95) node[cross,black] {};
    \draw [] (3.5,0.95) node[cross,black] {};
    \draw[thick, dotted](1.5, 0.2)--(4,0.2)--(4,2.2)--(1.5,2.2)--(1.5,0.2);
    \draw[] (2.75,2)--(2.75,2) node [] {\scriptsize$\rm {BSM}$};
   \draw[red, decorate,decoration={snake}, very thick] (0.5,1.2) to [out= 90, in = 90](5,1.2);
\end{tikzpicture}
\label{fig:remote}
\caption{Remote entanglement between two superconducting qubits. Each qubit-photon pair is composed of a transmon and a flying photon entangled system. Entanglement between different bosons is achieved by a CZ gate. BSM is performed with QNDs as described in the main text.  }
\label{fig:RemoteEntanglement}
\end{figure}
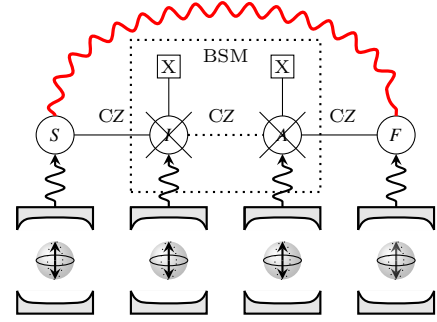
single photon counters, furthermore it is non-fault tolerant when amplification fails. 

Here we investigate utilizing our one-way NLA in order to perform the required BSM for remote entanglement. In Fig. (\ref{fig:RemoteEntanglement}) we consider an instance of our proposed protocol. Nodes $S$ \& $F$ are two remote nodes to be entangled, while $I$ \& $A$ are intermediate ones. At each node a transmon-cavity system is prepared in a qubit-photon entangled state, $\frac{1}{\sqrt{2}}(\lvert 0e \rangle + \lvert 1g\rangle)$ \cite{PhysRevLett.109.240501}. After that a photonic CZ gate entangles ; $SI$, $AF$, and $IA$ pairs respectively. The whole state of the system can then be written as 
    \begin{align}
    \lvert \varphi \rangle &=\Big[ \frac{1}{\sqrt{2}}(\lvert 0e \rangle_{S} + \lvert 1g\rangle_{S})\Big] \otimes \Big[ \frac{1}{\sqrt{2}}(\lvert 0e \rangle_{I} + \lvert 1g\rangle_{I})\Big] \nonumber \\ &\otimes \Big[ \frac{1}{\sqrt{2}}(\lvert 0e \rangle_{A} + \lvert 1g\rangle_{A})\Big] \otimes \Big[ \frac{1}{\sqrt{2}}(\lvert 0e \rangle_{F} + \lvert 1g\rangle_{F})\Big].
\end{align}
Applying a CZ between $SI$ and $AF$, transforms the state into
\begin{align}
    \lvert \varphi \rangle =&\Big[\frac{1}{2}  (\lvert 00ee \rangle_{SI} + \lvert 01eg \rangle_{SI} + \lvert 10ge\rangle_{SI} - \lvert 11gg \rangle_{SI})\Big] \otimes
   \nonumber \\  & \Big[\frac{1}{2}( \lvert 00ee \rangle_{AF} + \lvert 01eg \rangle_{AF} + \lvert 10ge\rangle_{AF} - \lvert 11gg \rangle_{AF})\Big].
\end{align}
Then another CZ is applied between nodes $I$ \& $A$
\begin{align}
   &\lvert \varphi \rangle = 
   \nonumber \\ &\frac{1}{4} \Big(\lvert 0++0eeee\rangle_{\tilde{S}\tilde{I}\tilde{A}\tilde{F}} + \lvert 0++1eeeg\rangle_{\tilde{S}\tilde{I}\tilde{A}\tilde{F}}\nonumber \\&+ \lvert 0+-0eege\rangle_{\tilde{S}\tilde{I}\tilde{A}\tilde{F}} -\lvert 0+-1eegg\rangle_{\tilde{S}\tilde{I}\tilde{A}\tilde{F}}\nonumber \\ &+ \lvert 0-+0egee\rangle_{\tilde{S}\tilde{I}\tilde{A}\tilde{F}} + \lvert 0-+1egeg\rangle_{\tilde{S}\tilde{I}\tilde{A}\tilde{F}} \nonumber \\ &- \lvert 0--0egge\rangle_{\tilde{S}\tilde{I}\tilde{A}\tilde{F}} +\lvert 0--1eggg\rangle_{\tilde{S}\tilde{I}\tilde{A}\tilde{F}} \nonumber \\ &+\lvert 1++0geee\rangle_{\tilde{S}\tilde{I}\tilde{A}\tilde{F}} +\lvert 1++1geeg\rangle_{SIAF} \nonumber \\ &+\lvert 1+-0gege\rangle_{SIAF} - \lvert 1+-1gegg\rangle_{SIAF} \nonumber \\ &- \lvert 1-+0ggee\rangle_{\tilde{S}\tilde{I}\tilde{A}\tilde{F}}-  \lvert 1-+1ggeg\rangle_{\tilde{S}\tilde{I}\tilde{A}\tilde{F}}  \nonumber \\ &+ \lvert 1--0ggge\rangle_{\tilde{S}\tilde{I}\tilde{A}\tilde{F}}- \lvert 1--1gggg\rangle_{\tilde{S}\tilde{I}\tilde{A}\tilde{F}}\Big).
\end{align}
Finally performing two X-basis measurements on the middle nodes completes the protocol. When the measurement outcomes are the same, we get 
\begin{align}
    \lvert \varphi^{\rm{(out)}}_{+} \rangle &= \frac{1}{\sqrt{2}} \big( \lvert 01eg \rangle_{\tilde{S}\tilde{F}} + \lvert 10 ge \rangle_{\tilde{S}\tilde{F}}\big).
    \label{eq:ES1}
\end{align}
On the other hand, when the outcomes are different, the state becomes  
\begin{align}
    \lvert \varphi^{\rm{(out)}}_{-} \rangle &= \frac{1}{\sqrt{2}} \big( \lvert 00ee \rangle_{\tilde{S}\tilde{F}} + \lvert 11 gg \rangle_{\tilde{S}\tilde{F}}\big).
    \label{eq:ES2}
\end{align}
By applying a Pauli-X correction to the second mode on both the photonic and atomic degrees of freedom, we get the same output as in Eq. (\ref{eq:ES1}). 

The previous protocol is a specific realization of the more general one-way NLA described earlier. When $\gamma =1$, and $t=1/2$ a one-way NLA reduces to the aforementioned remote entanglement swapping.

From Fig. (\ref{fig:NLAPsuccess}) we can directly estimate the success probability of the protocol. When detection efficiency is $\approx 0.85$, a one-way NLA based remote entanglement has an average success probability of $\approx 0.7$, whereas that of a QS based protocol is $\approx 0.45$. The success probability drops significantly when detection efficiency is further reduced. For the case of $\approx 0.5$ efficiency, a one-way NLA based protocol has on the average a probability of success of $\approx 0.38$, such that, OP1 success probability is $\approx 0.26$, while OP2's is $\approx 0.5$. On the other hand the success probability of a QS is $\approx 0.36$ which is clearly outperformed by OP2.   

Remote entanglement is the basis of many QKD protocols \cite{pirandola2020advances}. Many practical imperfections play a role in limiting the distance over which remote entanglement can be generated. The primary causes for the case at hand are qubit decoherence and propagation losses. For long distance entanglement sharing the qubit state has to be preserved for  milliseconds. Most recently, significant progress has been made towards longer qubit coherence times. Approximately the coherence time that can be achieved with the current state of the art technology is in the range of $1-3 \hspace*{0.2em}\text{ms}$ \cite{PhysRevLett.130.267001,place2021new}, which suggests a maximum sharing distance of $ \approx 30 \hspace*{0.2em}\text{km}$. As for attenuation due to propagation losses, %the absorption coefficient of the commercially available microwave waveguides inside a cryogenic environment is approximately $\approx 0.6 \hspace*{0.2em} \text{dB/km}$ \cite{meinhardt2018characterizing}. This corresponds to an attenuation length of $L_{\text{att}}\approx 7.3 \text{km}$, hence from Beer's law \cite{beer1852bestimmung},%
the survivability of a microwave wavepacket in a transmission medium decreases exponentially as a function of its length $L_{0}$. This is captured by the famous Beer's formula,\cite{beer1852bestimmung} $P(L_{0}) = \exp{(-L_{0}/2L_{\mathrm{att}})}$, where $L_{\mathrm{att}}$ is the attenuation length. 

In the next section we study in details the effect of transmission losses on SKR. More specifically, we consider two remote parties establishing a shared secure key by generating remote entanglement over a lossy channel. As before, we compare our one-way NLA entanglement sharing protocol to a QS's, and demonstrate the conditions under which our protocol outperforms both direct transmission and a QS. 
\subsection{Remote secret key generation}
\begin{figure*}
    \centering
   \subfloat[]{\includegraphics[scale=0.183]{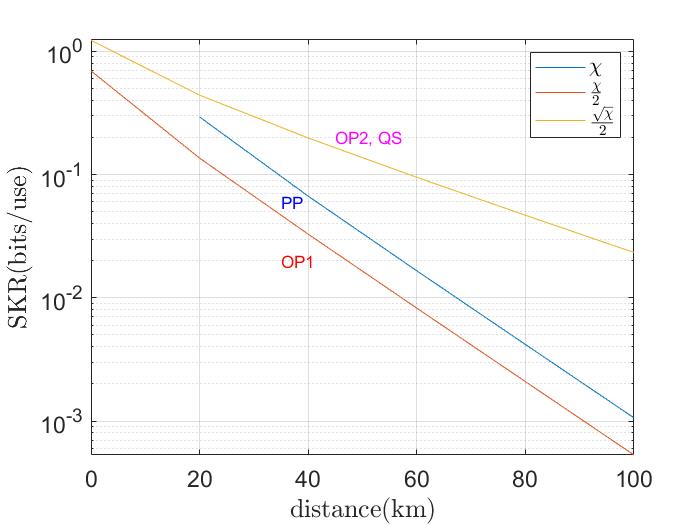}}
   \subfloat[]{\includegraphics[scale=0.183]{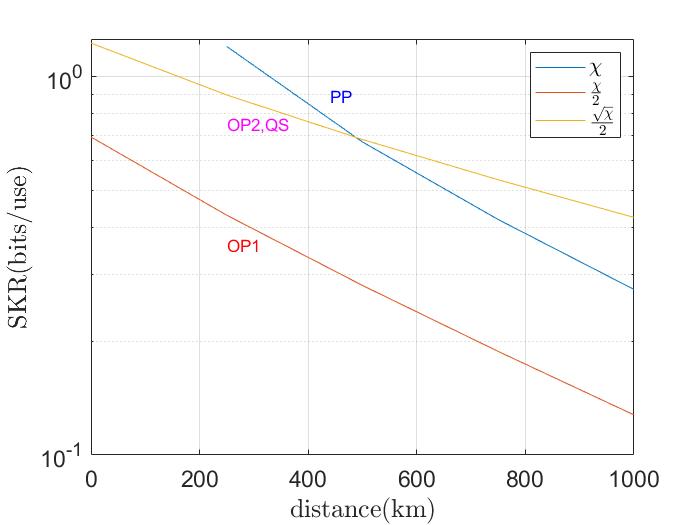}}
   \subfloat[]{\includegraphics[scale=0.183]{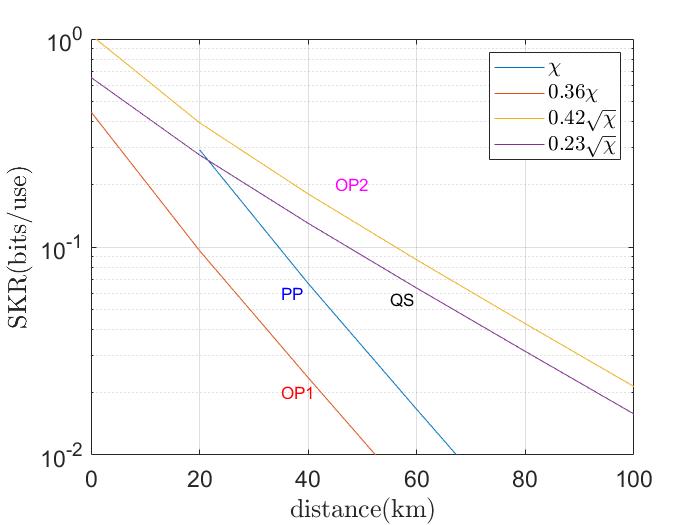}}
   \subfloat[]{\includegraphics[scale=0.183]{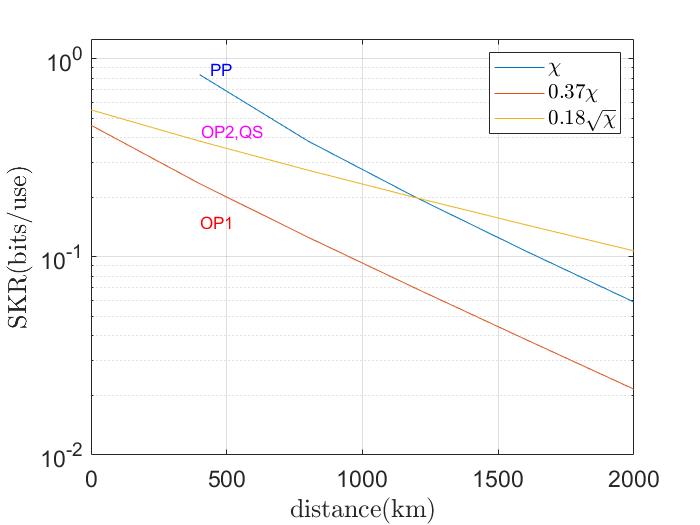}}
    \caption{SKR of our one-way NLA based on weak TMSV and QND detection. The plots depict SKR in (bits/use) against transmission distance in (km). Included in all plots the performance of a QS-NLA for comparison. All detectors in the first two plots are assumed ideal. It can be seen that OP2 and QS-NLA have the best performance in Fig. (6a), however, they both outperform a point to point link only for distances greater than $\approx 500$ $\mathrm{km}$ as shown in Fig. (6b). In Fig. (6c) detector's efficiency is assumed $\eta \approx 0.85$, and a dark count probability of $\mu=0.015$ for both one-way NLAs and $\mu=0.5$ for QS-NLA. In Fig. (6d) both detector's efficiency and dark count probability were assumed $0.5$ for all detectors. OP2 outperforms QS-NLA in Fig. (6c) and achieves the best performance, whereas OP1 has the least SKR. For a wireless transmission link, Fig. (6d) shows that OP2 \& QS-NLA has the highest SKR only for distances over 1200 $\mathrm{km}$. As before, OP1 has the least SKR. }
    \label{fig:SKR}
\end{figure*}
Consider now a slight variation of the previous protocol where the $S$ \& $I$ nodes (see Fig. (\ref{fig:RemoteEntanglement})) are prepared in a weak TMSV state identical to that in Eq. (\ref{eq:WeakTMSV}), while $A$ and $F$ are prepared in the same resource state described before, $\lvert \Phi \rangle_{AF} = \sqrt{1-t} \lvert 00 \rangle_{AF}+\sqrt{t}\lvert 11 \rangle_{AF }$. Suppose further, that modes $I$ and $A$ undergo two ADCs with transmissivities $\gamma$, and $\tilde{\gamma}$ respectively. 

Practically in such scenario each of the input and auxiliary photons are subjected to damping quantified by the aforementioned tranmissivities. After applying a CZ between nodes $I$ and $A$, we can write an expression for the overall state by including the two environment lossy modes as follows
\begin{align}
    \lvert \varphi \rangle &= \alpha \sqrt{1-t} \lvert 0+0+00 \rangle_{\tilde{S}\tilde{I}\tilde{E}\tilde{A}\tilde{F}\tilde{E'}} \nonumber \\ 
    &+\alpha \sqrt{t\tilde{\gamma}} \lvert 0+0-10 \rangle_{\tilde{S}\tilde{I}\tilde{E}\tilde{A}\tilde{F}\tilde{E'}} \nonumber \\
    &+ \alpha \sqrt{t(1-\tilde{\gamma})} \lvert 0+0+11 \rangle_{\tilde{S}\tilde{I}\tilde{E}\tilde{A}\tilde{F}\tilde{E'}} \nonumber \\ 
    &+ \beta \sqrt{\gamma (1-t)} \lvert 1-0+00 \rangle_{\tilde{S}\tilde{I}\tilde{E}\tilde{A}\tilde{F}\tilde{E'}} \nonumber \\ 
    &-\beta \sqrt{\gamma \tilde{\gamma}t} \lvert 1-0-10 \rangle_{\tilde{S}\tilde{I}\tilde{E}\tilde{A}\tilde{F}\tilde{E'}} \nonumber \\ 
    &+ \beta \sqrt{\gamma (1-\tilde{\gamma})t} \lvert 1-0+11 \rangle_{\tilde{S}\tilde{I}\tilde{E}\tilde{A}\tilde{F}\tilde{E'}} \nonumber \\ 
    &+ \beta \sqrt{(1-\gamma)(1-t)} \lvert 1+1+00 \rangle_{\tilde{S}\tilde{I}\tilde{E}\tilde{A}\tilde{F}\tilde{E'}} \nonumber \\ 
    &+ \beta \sqrt{t(1-\gamma) \tilde{\gamma}} \lvert 1+1-10 \rangle_{\tilde{S}\tilde{I}\tilde{E}\tilde{A}\tilde{F}\tilde{E'}} \nonumber \\ 
    &+ \beta \sqrt{t(1-\gamma)(1-\tilde{\gamma})} \lvert 1+1+11 \rangle_{\tilde{S}\tilde{I}\tilde{E}\tilde{A}\tilde{F}\tilde{E'}},
    \label{eq:EntSKR}
\end{align}
where $\tilde{E}$, $\tilde{E'}$ are two environment modes each in a vacuum state, and nodes $\tilde{I}$ and $\tilde{A}$ are written in the diagonal basis. 

After performing two X-basis measurements on nodes $\tilde{I}$, and $\tilde{A}$, the state becomes the following when the measurement outcomes are the same 
\begin{align}
    \lvert \varphi \rangle &=\frac{1}{\sqrt{N_{+}}} \Big[ (\alpha \sqrt{1-t} \lvert 00 \rangle_{\tilde{S}\tilde{F}} +\beta \sqrt{\gamma \tilde{\gamma}t} \lvert 11 \rangle_{\tilde{S}\tilde{F}}) \otimes \lvert 00 \rangle_{\tilde{E}\tilde{E'}} \nonumber \\ &+  \alpha \sqrt{t(1-\tilde{\gamma})} \lvert 01 \rangle_{\tilde{S}\tilde{F}} \otimes \lvert 01 \rangle_{\tilde{E}\tilde{E'}}  \nonumber \\ &+\beta \sqrt{(1-\gamma)(1-t)} \lvert 10 \rangle_{\tilde{S}\tilde{F}} \otimes \lvert 10 \rangle_{\tilde{E}\tilde{E'}}\nonumber \\ &+ \beta \sqrt{t(1-\gamma)(1-\tilde{\gamma})} \lvert 11 \rangle_{\tilde{S}\tilde{F}} \otimes \lvert 11 \rangle_{\tilde{E}\tilde{E'}}   \Big] .
    \label{eq:SKRSAME}
\end{align}
However, when the outcomes are different, it becomes
\begin{align}
    \lvert \varphi \rangle ¨&= \frac{1}{\sqrt{N_{-}}} \Big[ (\alpha \sqrt{t \tilde{\gamma}} \lvert 01 \rangle_{\tilde{S}\tilde{F}}+\beta \sqrt{\gamma(1-t)} \lvert 10 \rangle_{\tilde{S}\tilde{F}}) \otimes \lvert 00 \rangle_{\tilde{E}\tilde{E'}} \nonumber \\ &+ \beta \sqrt{\gamma t (1-\tilde{\gamma})} \lvert 11 \rangle_{\tilde{S}\tilde{F}} \otimes \lvert 01 \rangle_{\tilde{E}\tilde{E'}} \nonumber \\ &+ \beta \sqrt{\tilde{\gamma} t(1-\gamma)} \lvert 11 \rangle_{\tilde{S}\tilde{F}} \otimes \lvert 10 \rangle_{\tilde{E}\tilde{E'}} \Big] .
\end{align}
Then after applying a bit-flip operation on the final mode, $\tilde{F}$, the state becomes 
\begin{align}
    \lvert \varphi \rangle ¨&= \frac{1}{\sqrt{N_{-}}} \Big[ (\alpha \sqrt{t \tilde{\gamma}} \lvert 00 \rangle_{\tilde{S}\tilde{F}}+\beta \sqrt{\gamma(1-t)} \lvert 11 \rangle_{\tilde{S}\tilde{F}}) \otimes \lvert 00 \rangle_{\tilde{E}\tilde{E'}} \nonumber \\ &+ \beta \sqrt{\gamma t (1-\tilde{\gamma})} \lvert 10 \rangle_{\tilde{S}\tilde{F}} \otimes \lvert 01 \rangle_{\tilde{E}\tilde{E'}} \nonumber \\ &+ \beta \sqrt{\tilde{\gamma} t (1-\gamma)} \lvert 10 \rangle_{\tilde{S}\tilde{F}} \otimes \lvert 10 \rangle_{\tilde{E}\tilde{E'}} \Big],
    \label{eq:SKRDIFF}
\end{align}
where $N_{+} = \alpha^{2}(1-t\tilde{\gamma})+\beta^{2}[t \tilde{\gamma}(2 \gamma-1)+1-\gamma]$, and $N_{-}=\alpha^{2}t\tilde{\gamma}+\beta^{2}[\gamma +\tilde{\gamma}t(1-2\gamma)]$ are normalization constants.

Let us now analyze the previous equations in more details. In Eq. (\ref{eq:SKRSAME}) the first term on the right side represents an entangled state established between the remote parties when each environment mode is in a vacuum state. The rest of the terms are error states corresponding to the following loss events respectively: when the second environment mode $\tilde{E'}$ captures a photon, loss occurs to the auxiliary mode, whereas when $\tilde{E}$ captures a photon, it occurs to the input. The final term means that both photons were lost. Similarly, after error correction, the terms in Eq. (\ref{eq:SKRDIFF}) corresponds to the following: the first term represents a successful entangled state as can be verified by inspecting the environment vacuum states, whereas the rest of the error terms correspond to a loss in either the input or the auxiliary modes depending on which environment mode captured a photon. Unlike the state in Eq. (\ref{eq:SKRSAME}), the event where both the input and auxiliary modes are lost doesn't occur, since the physical meaning of different measurement outcomes corresponds to detecting one and only one photon in either $\tilde{I}$ or $\tilde{A}$. 

Focusing now on SKG,  the repeater-less bound \cite{pirandola2017fundamental} for  two remotely communicating parties defines a relation between the SKR that can be generated and the tranmsissivity of the linking channel, ${\rm {SKR}} \approx - {\rm {log}}_{2}(1-\chi^{1/K})$, where $\chi$ is the channel's transmissivity, and $K$ is the number of repeater links. When $K=1$, SKR corresponds to a point-to-point direct transmission link, whereas when $K=2$ it corresponds to a single repeater case.

The inner workings of one-way NLA resembles that of a quantum repeater, this can be shown by noting that the device's BSM swaps the entanglement between the middle nodes to the remote ones. Furthermore, a secret key rate (SKR) can be estimated according to the success probability of the device \cite{winnel2021overcoming}, since it establishes an entanglement channel between the involved remote parties. 

When the outcomes of a BSM are identical, i.e., OP1, the success probability is $P^{\mathrm {succ}}_{S} \approx \alpha^{2}(1-t)+\beta^{2}\gamma \tilde{\gamma}t$, whereas when they are  different, OP2, it becomes $P^{\mathrm {succ}}_{\mathrm{D}} \approx \alpha^{2}t \tilde{\gamma}+\beta^{2} \gamma (1-t)$. In the limit of small squeezing $\lambda < 1$, $\alpha^{2} \approx 1$, $\beta^{2} \approx \lambda$, symmetric loss, $\gamma = \tilde{\gamma} = \sqrt{\chi}$, $t=1/2$, it can be seen that the success probabilities become $P_{S}^{\mathrm {succ}} \propto \chi/2$, and $P^{\mathrm {succ}}_{\mathrm{D}} \propto \sqrt{\chi} / 2$ respectively. The previous approximations can be justified by observing that the first term in both success probability expressions is a vacuum offset, since it corresponds to the event were both the input and the auxiliary modes are in vacuum states.  

In Fig. (\ref{fig:SKR}) we have studied the behaviour of SKR (in bits/use) against the transmission distance (in km) for two different transmission media and both ideal and non-ideal detection. We have included in this analysis SKR of a point to point link as an indicative of the PLOB bound. The first two panels study the case where the device's detectors are operating in the ideal limit, whereas the last two plots consider detection imperfections. For the ideal case, the performance of a one-way NLA OP2 and a QS-NLA coincide.

We have taken in Fig. (\ref{fig:SKR}a) a cryogenic waveguide transmission medium with 0.6 $\rm {dB/km}$ attenuation factor, whereas in Fig. (\ref{fig:SKR}b) a wireless transmission link with $0.0063 {\rm {dB/km}}$ attenuation factor \cite{fesquet2023perspectives} was assumed. In the former, from Beer's law, the attenuation length was found to be $L_{\rm {att}}=7.3 \hspace*{0.2em}\rm {km}$, whereas in the latter it was $L_{\rm {att}} \approx 700 \hspace*{0.2em}\rm {km}$. The last two plots consider the same transmission media, however, detection non idealities were considered. Similar to Fig. (\ref{fig:SKR}a), Fig.  (\ref{fig:SKR}c) considers the same channel attenuation factor, while detector's efficiency is $\eta  \approx 0.85 $. Furthermore, we have assumed dark count probability of $\mu \approx 0.015$ for one-way NLA and $\mu \approx 0.5$ for QS-NLA, since microwave PNRD are currently unavailable as mentioned earlier. On the other hand, in Fig. (\ref{fig:SKR}d) a wireless transmission medium was assumed as in Fig. (\ref{fig:SKR}b), however,  both detector's efficiency and dark count probability were assumed $0.5$ for all detectors.

In Fig. (\ref{fig:SKR}a), it can be seen that one-way NLA operating under OP2 and QS-NLA have the best SKR. This can be justified by noticing that when the channel's transmissivity is small, splitting the overall distance by a NLA outperforms direct transmission, due to its entanglement properties. Comparatively a higher channel transmissivty as in Fig. (\ref{fig:SKR}b), namely $0.0063 \hspace*{0.2em} \rm {dB/km} $, resulted in an improved performance over direct transmission only for large distances, $L_{0} > 500 \rm {km}$. We note that in both of the previous cases when one-way NLA operated under OP1 it had the least SKR. 

When detection imperfections were considered, SKR of all protocols dropped as shown in Fig. (\ref{fig:SKR}c), however, a practical QS-NLA would be outperformed by a one-way NLA operating under OP2, due to the unavailability of perfect microwave PNRD. Similarly as before one-way NLA OP1 has the least performance. For the last case where the detection limits of all detectors are the same, Fig. (\ref{fig:SKR}d) showed that direct transmission had the best SKR among all channels for distances less than $\approx 1200 \rm {km}$, nonetheless, when the distance exceeds this limit, it can be seen that one-way NLA OP2 \& QS-NLA have the best SKR, thus suggesting that they are more suitable for long distance microwave QIP.  

To summarize, in this section we have shown that due to the inherent resemblance between one-way NLA and a quantum repeater, our device can be utilised to establish a remote entanglement channel for SKG. The device's success probability translates to a SKR that can be generated. Furthermore, it can beat the repeater-less bound when operating under OP2 and outperform a QS-NLA for lossy cryogenic links, thus offering a more secure option for cryogenic chip-to-chip communications \cite{wallraff2018deterministic}.   
\section{CONCLUSION}
In this paper we have proposed a novel microwave bosonic NLA based on a simple one dimensional linear cluster of four nodes. Unlike conventional QS-NLAs, our proposed device doesn't rely on PNRDs to signal the success of the BSM required for amplification. Instead, our device utilizes QND detectors operating in the photodetection limit, which significantly eases the experimental requirements for a microwave NLA. Operationally, the proposed device inherits the fault tolerant capabilities of a cluster state, which renders it capable of correcting the erroneous output and achieve amplification. This feature is not possible in a conventional QS-NLA, which has to repeat the whole process when amplification fails.

Owing to the low dark count probability of a QND detector, our device showed an enhanced probability of success over a QS-NLA operating with microwave PNRDs. Furthermore, our device also demonstrated a great ability to enhance microwave QIP tasks. For quantum sensing applications, we have shown that with the aid of a one-way NLA, entanglement enhanced sensing protocols can fully restore an attenuated idler beam despite detection imperfections. This was achieved at the expense of a fewer increase in the number of probe pairs, which is still significantly lower than that required by a QS-NLA performing the same task. Furthermore, we have also shown that our device can generally establish secret keys between remote parties at a rate surpassing that of a direct communication link, both for a cryogenic ADC and a long distance wireless channel. More specifically, it has showed higher SKR than that of a QS-NLA for cryogenic ADC when detection inefficiencies were considered. 

Finally we note that the proposed device model can be also employed in the optical domain, thus rendering it a new addition to the QIP toolbox.    

\EOD
\end{document}